%% file: 00_main.tex
\documentclass[conference, 10pt]{IEEEtran}
\IEEEoverridecommandlockouts

\usepackage[printonlyused, nohyperlinks]{acronym}
\usepackage{booktabs}
\usepackage{siunitx}
\usepackage[english]{babel}

\usepackage{tikz}
\usepackage{algorithmic}
\usepackage[ruled,vlined]{algorithm2e}%

\SetCommentSty{mycommfont}
\SetAlCapNameFnt{\small}
\SetAlCapFnt{\small}
\usepackage{threeparttable}

\usepackage{pgfplots}
\usepackage{pgfplotstable}
\usepgfplotslibrary{patchplots, colormaps, groupplots, dateplot}
\pgfplotsset{compat=1.18}
\usetikzlibrary{arrows, chains, positioning,fit,calc, shapes, shapes.geometric, matrix, decorations.pathmorphing, automata, shapes.gates.logic.US,shapes.gates.logic.IEC,calc, external, decorations.pathreplacing, spy}
\usepackage{ifthen, xstring, calc, pgfopts}

\usepackage{xstring}
\usepackage{calc}
\usepackage{pgfopts}
\usepackage{mathtools}%
\usepackage{tabularx, booktabs}
\newcolumntype{Y}{>{\centering\arraybackslash}X}
\newcolumntype{Z}{>{\raggedleft\arraybackslash}X}
\newcolumntype{C}[1]{>{\centering\arraybackslash}p{#1}}
\usepackage{rotating}
\usepackage{multirow}
\usepackage{graphicx}
\usepackage{subfig}
\usepackage{textcomp}
\usepackage[shortlabels]{enumitem}

\usepackage[english]{babel}
\usepackage[cmintegrals]{newtxmath}

\definecolor{hsurot}	{cmyk}{.00 1 .59 .26}
\definecolor{hsugrau}	{cmyk}{.38 .37 .39 .15}
\definecolor{hsugelb}	{cmyk}{0 .16 .80 0}
\definecolor{hsublau}	{cmyk}{1 .40 0 .82}
\definecolor{hsuturkis}	{cmyk}{1 .14 .60 .49}
\definecolor{hsugruen}	{cmyk}{.16 .16 .91 .28}
\definecolor{hsubraun}	{cmyk}{.00 .57 1 .17}
\definecolor{hsuorange}	{cmyk}{.01 .87 .77 .13}

\usepackage[numbers, sort&compress]{natbib}
\usepackage[final]{microtype}
\usepackage[hidelinks]{hyperref}
\usepackage{xurl} 

\newcommand{\acrshort}[1]{\acs{#1}}

\begin{document}
\emergencystretch=1em
\title{Coupling Scenario-Based Grid Simulations with State Estimation: Measurement Requirements for Low-Voltage Networks under the German Energy Transition Pathway%
\thanks{This work is part of the project DISEGO which was supported by the German Federal Ministry for Economic Affairs and Energy and the Projektträger Jülich GmbH (FKZ: 03EI6078A/03EI6078H). This work was also partially carried out within the project OptiFlex which is funded by dtec.bw -- Digitalization and Technology Research Center of the Bundeswehr.}
}

\author{
\IEEEauthorblockN{Nane Zimmermann$^{1}$, Lukas Peter Wagner$^{1,*}$, Luca von R\"onn$^{1}$,\\
Florian Strobel$^{2}$, Paul H\"uttmann$^{2}$, Felix Gehlhoff$^{1,*}$}
\IEEEauthorblockA{$^{1}$\textit{Institute of Automation Technology},
Helmut Schmidt University, Hamburg, Germany}
\IEEEauthorblockA{$^{2}$\textit{PSI Software SE}, Berlin, Germany}
\IEEEauthorblockA{$^{*}$Correspondence: lukas.wagner@hsu.hamburg~(L.P.W.),\\
felix.gehlhoff@hsu.hamburg~(F.G.)}
}

\hfuzz=6pt
\maketitle
\hfuzz=0.1pt

\begin{abstract}
Increasing penetration of electric vehicles, heat pumps, and rooftop photovoltaics is creating thermal and voltage stress in low-voltage distribution grids. This work links the German Federal Government energy transition pathway (2025--2045) with state estimation performance requirements, evaluated at five milestone years from 2025 to 2045 on two SimBench reference networks across three equipment quality levels (good, medium, poor) and three VDE Forum Netztechnik/Netzbetrieb (VDE~FNN) measurement constellations that differ in the availability of transformer and feeder-level instrumentation. Within this work's analysis, congestion is caused exclusively by transformer overloading and voltage-band violations. No individual line exceeds its thermal rating (maximum: 98.6\%). Equipment quality governs congestion onset for a given deployment trajectory: under good equipment, congestion remains absent through 2045, under medium equipment it emerges from 2035 (4 of 10 scenarios), under poor equipment from 2025 (9 of 10). Without transformer instrumentation, median voltage estimation errors reach 6--42\% regardless of smart meter penetration. Adding a single transformer measurement reduces errors by an order of magnitude, achieving median errors of 0.5--1.4\%. In urban networks, transformer-level instrumentation meets the VDE~FNN voltage accuracy target (99th percentile voltage error below 2\%) in all configurations. In rural networks under poor equipment, the target is approached but not met. These findings motivate prioritizing transformer instrumentation as an effective first step for grid observability and supplementing the current consumption-driven metering rollout with risk-based deployment criteria linked to local congestion exposure.
\end{abstract}

\renewcommand{\sectionautorefname}{Section}
\renewcommand{\subsectionautorefname}{Section}
\renewcommand{\subsubsectionautorefname}{Section}
\renewcommand{\figureautorefname}{Figure}
\newcommand{\subfigureautorefname}{Figure}
\newcommand{\pgfplotslinkautorefname}{\figureautorefname}
\renewcommand{\equationautorefname}{Equation}
\renewcommand{\algorithmautorefname}{Algorithm}

\newcommand{\requ}[1]{\protect\begin{tikzpicture}[baseline=-0.35em]%
\protect\draw[black] (0,0) circle [inner sep=0pt, outer sep=0pt, radius=0.17] node {\scriptsize \textbf{#1}};%
\end{tikzpicture}}
\newcommand*\emptycirc[1][1ex]{\tikz\draw[thick] (0,0) circle (#1);}
\newcommand*\halfcirc[1][1ex]{%
  \begin{tikzpicture}
  \draw[fill] (0,0)-- (90:#1) arc (90:270:#1) -- cycle ;
  \draw[thick] (0,0) circle (#1);
  \end{tikzpicture}}
\newcommand*\fullcirc[1][1ex]{%
  \begin{tikzpicture}
  \draw[thick, fill] (0,0) circle (#1);
  \end{tikzpicture}}
\newcommand*\notapplicable[1][1.75ex]{\tikz[baseline=-0.1em]\draw[fill, hsugrau] (0,0) rectangle (#1,#1);}

\input{1_introduction}
\input{2_related_work}
\input{3_mathematical_analysis_state_estimation}
\input{4_development_pathways_load_limits}
\input{5_state_estimation_quality}
\input{7_discussion}
\input{8_summary_outlook}
\setcounter{section}{0}
\appendix \section{Supplementary Information} \label{appendixsection}

\subsection{Algorithm Mirror Test Results}\label{subsec:mirror_test_appendix}

\autoref{tab:mirror_test_results} reports the normalised mean voltage deviation for each of the ten scenarios under complete real measurement coverage (mirror test), evaluated over the first three congestion periods per scenario. The acceptance criterion is a deviation below $0.1\,\%$. All scenarios pass. The 2035, 2040, and 2045 rural scenarios exhibit slightly elevated deviations (up to $0.09\,\%$) that remain well within the acceptance criterion and reflect residual numerical effects of the \ac{WLS} solution rather than a systematic bias (see \autoref{subsec:algorithm_verification_se}).

\begin{table}[htbp]
\centering
\caption{Mirror test: normalised mean voltage deviation per scenario (mean over three periods).
         Criterion: ${<}0.1\,\%$. All scenarios pass.}
\label{tab:mirror_test_results}
\footnotesize
\setlength{\tabcolsep}{4pt}
\renewcommand{\arraystretch}{1.2}
\begin{tabularx}{\columnwidth}{l >{\centering\arraybackslash}X >{\centering\arraybackslash}X >{\centering\arraybackslash}X >{\centering\arraybackslash}X}
\noalign{\hrule height 1.1pt}
\textbf{Scenario} & \textbf{Period 1} & \textbf{Period 2} & \textbf{Period 3} & \textbf{Status} \\
\noalign{\hrule height 0.8pt}
2025 rural      & $<$0.05\,\% & $<$0.05\,\% & $<$0.05\,\% & \checkmark \\
2025 urban      & $<$0.05\,\% & $<$0.05\,\% & $<$0.05\,\% & \checkmark \\
\noalign{\hrule height 0.4pt}
2030 rural      & $<$0.05\,\% & $<$0.05\,\% & $<$0.05\,\% & \checkmark \\
2030 urban      & $<$0.05\,\% & $<$0.05\,\% & $<$0.05\,\% & \checkmark \\
\noalign{\hrule height 0.4pt}
2035 rural      & 0.06\,\%    & 0.05\,\%    & $<$0.05\,\% & \checkmark \\
2035 urban      & $<$0.05\,\% & $<$0.05\,\% & $<$0.05\,\% & \checkmark \\
\noalign{\hrule height 0.4pt}
2040 rural      & 0.07\,\%    & 0.06\,\%    & 0.05\,\%    & \checkmark \\
2040 urban      & $<$0.05\,\% & $<$0.05\,\% & $<$0.05\,\% & \checkmark \\
\noalign{\hrule height 0.4pt}
2045 rural      & 0.09\,\%    & 0.08\,\%    & 0.06\,\%    & \checkmark \\
2045 urban      & $<$0.05\,\% & $<$0.05\,\% & $<$0.05\,\% & \checkmark \\
\noalign{\hrule height 1.1pt}
\end{tabularx}
\end{table}

\bibliographystyle{IEEEtranN}
\section*{Acknowledgments}
The authors thank the Technical University Hamburg for providing their implementation of the state estimation algorithm \cite{strobelUncertaintyQuantificationBranchCurrent2024}.

\section*{Abbreviations}
The following abbreviations are used in this work:
\begin{acronym}[VDE~FNN]
\input{acronyms}
\end{acronym}

\bibliography{literature}
\end{document}

%% file: 1_introduction.tex
\section{Introduction} \label{sec:introduction}
The transformation of energy systems toward climate neutrality is driven by the extensive electrification of end-use sectors and the large-scale integration of \acp{DER} such as rooftop photovoltaics, \acp{EV}, and heat pumps \cite{Khan2026}. Because these assets are primarily integrated at the \ac{LV} level, a grid tier historically characterized by limited monitoring infrastructure and a design centered on unidirectional power flows, their proliferation necessitates a fundamental shift in operational management \cite{Pham2025}. In this context, Germany represents a particularly relevant case study due to its advanced energy transition strategy and its legally binding target of achieving climate neutrality by 2045. The German regulatory framework § 14a \ac{EnWG} mandates a paradigm shift from \emph{grid follows load} to \emph{load follows grid}, empowering \acp{DSO} to employ curative control through the temporary curtailment of controllable local systems during periods of acute congestion. While this regulatory flexibility facilitates a more efficient utilization of existing grid capacity compared to conservative static planning, it remains strictly contingent upon meeting advanced requirements for grid observability and \ac{SE} \cite{Para14a}. Consequently, these developments make Germany a suitable environment for analyzing the evolving operational and monitoring requirements of increasingly decentralized and bidirectional distribution networks.

Unlike transmission grids, \ac{LV} distribution networks were originally designed for passive operation with unidirectional power flows and limited simultaneity of high-power loads. The increasing penetration of \ac{DER}, combined with flexible and electrified demand, leads to higher transformer and line loading as well as voltage band violations \cite{kilthauDecentralizedOptimizationApproach2025, kilthauGeneralizedDistributedEnergy2025}. Reinforcement of distribution infrastructure is capital-intensive and slow, making operational congestion management and improved monitoring essential in the medium term \cite{Mehmood2026,Pham2025}.

Scenario-based studies frequently analyze alternative energy transition pathways for \ac{DER} expansion and quantify their impact on congestion frequency, hosting capacity, or flexibility requirements. However, many of these analyses implicitly assume full observability of the network state \cite{Zhang2024,Falabretti2018}. In practice, \ac{LV} grids suffer from sparse and heterogeneous measurement coverage, asynchronous data streams, and incomplete knowledge about topology and parameters \cite{wagnerSystematicReviewState2026}. This persistent observability gap fundamentally limits the reliable detection and, thus, assessment of emerging congestion situations.

As a key enabler of distribution grid monitoring, \ac{SE} has therefore gained increasing attention. \ac{LV} \ac{SE} approaches typically combine transformer measurements, smart meter data, and pseudo-measurements derived from advanced load profiling of historical consumption data. Many implementations rely on \ac{WLS}-based formulations. However, the demand for higher observability in increasingly active grids highlights the inherent challenges of \ac{LV} \ac{SE} \cite{Abdel-Majeed2013,Taczi2021,Azzopardi2024}. While these methods can significantly improve situational awareness, their accuracy strongly depends on the measurement configuration and the statistical properties of pseudo-measurements~\cite{strobelUncertaintyQuantificationBranchCurrent2024}.

At the same time, regulatory frameworks such as § 14a \ac{EnWG} in Germany introduce new requirements for grid-oriented control of flexible loads via smart meter infrastructures \cite{KarmannPragmaticInteroperability}. This regulation mandates agreements for the grid-oriented control of devices such as heat pumps and \ac{EV} charging points in exchange for reduced network charges. To ensure non-discriminatory operation, it empowers \acp{DSO} to manage these loads through economic incentives or direct power limits while simultaneously requiring more precise digital monitoring of the distribution network. These developments increase both the availability of measurement data and the operational need for reliable grid state assessment. In critical or non-routine situations, insufficient data quality or uncertainty in the estimated grid state may require manual intervention by \acp{DSO} \cite{wagnerSupportingManualDecisionMaking2025}.

Despite substantial progress in both scenario-based grid studies and \ac{LV} SE research, the interaction between long-term \ac{DER} energy transition pathways, resulting congestion patterns, and the measurement requirements for reliable congestion detection remains insufficiently quantified. In particular, it is unclear what measurement configuration (smart-meter penetration together with the availability of transformer- and feeder-level instrumentation) is required to detect congestion events with sufficient accuracy and regulatory compliance as electrification increases.

This work addresses this gap by linking the German Federal Government energy transition pathway for the period 2025--2045 with a systematic evaluation of \ac{SE} performance under the three \ac{VDEFNN} measurement constellations at the current regulatory minimum \ac{SMGW} penetration. Three research questions structure the investigation:

\begin{description}
  \item[RQ1:] Considering different grid equipment configurations, at what points in time do thermal and voltage limit violations emerge in representative rural and urban LV distribution grids, how do their frequency and severity evolve, and which grid components are primarily affected?
  \item[RQ2:] How does \ac{SE} accuracy and congestion detection capability depend on the type of measurement infrastructure (transformer, feeder, and smart meter instrumentation), and which constellation meets established quality targets for the congestion scenarios identified in RQ1?
  \item[RQ3:] What regulatory adjustments to smart meter rollout requirements and grid monitoring standards are necessary to ensure adequate grid observability under the expected \ac{DER} deployment trajectories, specifically in the context of § 14a~\ac{EnWG}?
\end{description}

To answer these questions, this work proposes a methodological framework that (i) derives congestion events and their characteristics from scenario-based time-series simulations, (ii) evaluates \ac{SE} accuracy and the detectability of thermal and voltage limits as a function of measurement concepts, and (iii) derives quantitative requirements for measurement infrastructure and monitoring with direct implications for current German regulations.

The main contributions of this work are:
\begin{enumerate}
  \item Coupling of the German energy transition pathway (2025--2045) to time-series simulations across three equipment quality levels, showing that equipment quality governs congestion onset and severity in rural and urban \ac{LV} reference networks.
  \item Accuracy assessment of \ac{BC-Mod} \ac{WLS} \ac{SE} under three \ac{VDEFNN} measurement constellations, showing that the measurement constellation, not smart-meter penetration, governs \ac{SE} accuracy: a single transformer measurement reduces median voltage errors by an order of magnitude compared to smart-meter-only configurations, regardless of penetration.
  \item Evidence that congestion is caused exclusively by transformer overloading and voltage-band violations, making voltage estimation the operationally relevant \ac{SE} contribution. The \ac{VDEFNN} voltage target ($f_V^{p99} \leq 2\,\%$) is met by K2 in urban networks.
  \item Regulatory recommendations: prioritized transformer instrumentation, risk-based \ac{SMGW} densification, and alignment of \ac{VDEFNN} detection metrics with congestion-relevant voltage thresholds.
\end{enumerate}

The remainder of this work is structured as follows: \autoref{sec:related_work} reviews related work. \autoref{sec:mathematical_analysis_state_estimation} defines the \ac{SE} metrics and algorithm. \autoref{sec:impacts_development_pathways_load_limits} analyzes the impacts of the energy transition on load limits in \ac{LV} grids. \autoref{sec:state_estimation_quality} addresses \ac{SE} quality under varying measurement availability. \autoref{sec:discussion} discusses the results, derives regulatory implications, and identifies limitations. \autoref{sec:conclusion} summarizes the work and outlines directions for future research.

%% file: 2_related_work.tex
\section{Related Work} \label{sec:related_work}
This work aims to quantify what monitoring infrastructure (the measurement configuration and density) is required to ensure reliable congestion detection via \ac{SE} as electrified demand and distributed generation grow in \ac{LV} grids. To position this contribution, prior work is assessed against the following derived requirements (R):

\begin{enumerate}[label=\textbf{R\arabic*:}, leftmargin=*, align=left]
  \item \textbf{Pathway-to-feeder downscaling.} The work must translate macro-level technology scenarios (e.g., heat pumps, \acp{EV}, rooftop \ac{PV}) into feeder-level load and generation profiling, including spatial uncertainty and simultaneity \cite{Becker2025,Fakhrooeian2024}.
  \item \textbf{Physically grounded congestion characterization.} The work must quantify voltage-band violations and thermal overloads using time-series or probabilistic grid simulations, relating outcomes to the actual topology and parameters of each feeder \cite{Fakhrooeian2024,Protopapadaki2017}.
  \item \textbf{Operational consequences and flexibility.} The work must either (i) explicitly model active operation (e.g., active network management, coordinated charging) or (ii) clearly distinguish them from passive \emph{fit-and-forget} scenarios, i.e., scenarios that assume no curtailment or flexibility activation to resolve constraints \cite{Damianakis2025,Delchambre2023,Haendel2023}.
  \item \textbf{Observability-aware monitoring.} \ac{SE} and monitoring methods for \ac{LV} grids must cope with sparse, heterogeneous, and potentially asynchronous measurements, including potential dependence on pseudo-measurements \cite{Paruta2021,Buason2024,wagnerSystematicReviewState2026}.
  \item \textbf{Measurement-to-detectability link.} Beyond \ac{SE} accuracy, the work must connect measurement penetration (and sensor types) to the \emph{detectability} of operational limit violations (congestion events) \cite{Buason2024,Dehbozorgi2025}.
  \item \textbf{Benchmarking and reproducibility.} Evaluation must be grounded on reproducible benchmark grids and time-series datasets (e.g., SimBench) to support systematic comparison of scenario results and \ac{SE} performance \cite{Fakhrooeian2024,Treutlein2026,Meinecke2020}.
\end{enumerate}

The following subsections synthesize related work along these requirements and derive the research gap. 

\subsection{Scenario Downscaling and Benchmarking} \label{subsec:category1}
Luo et al. \cite{Luo2026} propose an interpretable diffusion model to generate long-term distribution load scenarios. Their method integrates a transformer and temporal decomposition to capture multi-scale load patterns across seasons and years. The generated scenarios achieve high fidelity in temporal and frequency domains. However, the method operates on aggregate system load data and does not explicitly translate macro-level technology scenarios (heat pumps, \acp{EV}) to feeder profiles, so (R1) is only partially met. It also does not simulate grid constraints (R2) or incorporate passive operation, SE, or reproducible benchmark grids (R3--R6). Fakhrooeian et al. \cite{Fakhrooeian2024} systematically evaluate \ac{EV} charging and heat pump impacts on German \ac{LV} grids using 200 realistic feeder models. They create scenarios of \ac{EV} and heat pump penetration and run time-series power flows. The study quantifies minimum voltages and maximum transformer/line loadings for each scenario. Thus it directly addresses (R1) by downscaling national \ac{EV}/heat pump adoption and using a physical congestion analysis (R2) simultaneously \cite{Fakhrooeian2024}. For example, they observe voltage dips and cable overloads at high heat pump penetration. They use real feeder data (R6). They assume passive operation (R3) but do not tackle \ac{SE} or measurement issues (R4, R5). Overall, this paper provides rich feeder-level scenarios with stress analysis (R1--R3, R6), but leaves out monitoring and control aspects (R4, R5). Treutlein et al. \cite{Treutlein2026} release the FeederBW dataset, containing two years of measurements from 200 German LV feeders. This dataset includes one-minute power flows, weather, and detailed metadata (housing count, PV/EV/heat pump capacities). By offering real-world feeder profiles under evolving \ac{DER} adoption, it fully supports (R6). It is intended for applications like load forecasting, machine learning, and synthetic data generation. However, as a data publication it does not perform modeling or downscaling itself (R1--R5). One of its values lies in enabling future researchers to generate realistic feeder scenarios and test methods (R6). Meinecke et al. \cite{Meinecke2020} introduce SimBench, a benchmark suite of synthetic German power systems. SimBench provides \ac{LV}/\ac{MV}/\ac{HV} grid models and one-year time series of loads, \ac{PV}, and \ac{EV} profiles. The dataset explicitly aims to standardize test cases for research. SimBench “simplifies reproducing study results” by offering scenario data \cite{Meinecke2020} fulfilling (R6). SimBench’s future scenarios can serve as high-level inputs, but the work itself does not generate downscaling or analyze grid congestion (R1--R5).

\subsection{Grid Congestion Analysis and Operational Flexibility}\label{subsec:category2}
Damianakis et al. \cite{Damianakis2025} provide a thorough review of \ac{LV}-grid impacts of \acp{PV}, \acp{EV}, and heat pumps. They summarize studies showing that high \acp{DER} penetration can cause voltage rise violations, phase unbalance, power losses, and overloading. The review also highlights mitigation: violations can be significantly reduced through coordinated voltage/reactive power control and tap changer optimization, which adjusts the transformation ratio to stabilize the grid voltage. This work addresses grid effects (R2) and operational measures (R3), but as a survey it does not perform new downscaling (R1) or involve SE (R4--R5). Protopapadaki and Saelens \cite{Protopapadaki2017} simulate multiple LV feeders using stochastic building and load models to study PV and heat pumps effects. They use a Monte Carlo simulation with detailed, Modelica-based building models, which account for the thermal behavior of the building envelope, its orientation, and the window-to-wall ratio to capture realistic heat pump load profiles. Their results show that air-source heat pumps impose larger voltage drops and cable overloads than rooftop PV in the studied feeders. They find rural feeders overload at heat pump penetrations from 20\% to 30\%. This directly quantifies congestion (R2). No active control (R3) or SE is included (R4--R5). Delchambre et al. \cite{Delchambre2023} present a probabilistic power-flow analysis for a single LV feeder with battery assets. It covers physical congestion due to random frequency containment reserve and automatic frequency restoration reserve injections (R2), and the operational impact of these frequency services (R3). However, it has no scenario-downscaling (R1), no \ac{SE} (R4), and no detection strategy (R5), and it is not based on a public benchmark (R6). Khan et al. \cite{Khan2026} review optimization and control methods for \ac{LV} networks with high \ac{DER} share. They survey decentralized DER, demand response, and the regulation of voltage levels through reactive power management, typically provided by smart inverters. Notably, they summarize quantitative effects of control: coordinated voltage control and optimal tap settings can improve hosting capacity and reduce constraint breaches by 20--67\% \cite{Khan2026}. This covers active methods (R3) and indirectly less congestion (R2). However, it is a literature review (no new data) and does not perform scenario downscaling or SE (R1, R4, R5).
 
\subsection{Observability and Detectability}\label{subsec:category3}
Fotopoulou et al. \cite{Fotopoulou2022} review \ac{SE} algorithms under sparse measurements. They highlight that limited real-time data in \ac{LV} grids negatively affects \ac{SE} accuracy and convergence. This is tackled either with more instruments or with pseudo-measurements \cite{Fotopoulou2022}. The review covers various \ac{SE} formulations (\ac{WLS}, Kalman) and the use of pseudo-measurements to maintain observability. Thus it addresses observability (R4) and partially discusses pseudo-measurements for unobserved states (R5). It does not involve scenario downscaling or grid simulation (R1--R3, R6). Mattoo et al. \cite{Mattoo2025} present a modular two-level Kalman filter state estimator for distribution grids with asynchronous heterogeneous measurements. In Stage 1, separate estimators process micro-phasor measurement units, remote terminal units, and smart meter data. Stage 2 fuses these into a global estimate. They explicitly handle different refresh rates and latencies, enabling real-time SE with mixed sensors. The measurements from these different devices are asynchronous. This satisfies observability with sparse/heterogeneous data (R4). It does not explicitly address violation detection (R5) nor other requirements (R1--R3, R6). Paruta et al. \cite{Paruta2021} design a greedy algorithm for meter placement on \ac{LV} networks. They use an enhanced DistFlow \ac{SE} model (including cable capacitances) to evaluate observability. Iteratively they place one measurement device that improves SE accuracy the most, until predefined SE error limits are met. On a 75-node Swiss feeder, they achieve full observability with the minimum number of required measurement devices. This work directly addresses (R4) by improving \ac{SE} observability. It does not deal with violation detection (R5) or other requirements (R1--R3, R6). Buason et al. \cite{Buason2024} formulate a bilevel optimization to place a minimal number of voltage sensors and set alarm thresholds so that any violation of voltage limits will be detected. The upper-level minimizes sensor count and false alarms, while the lower-level computes maximum voltages given uncertain injections. They use linear power-flow approximations to guarantee no violation escapes detection. Thus it explicitly ensures detectability of limit violations (R5) and optimizes sensor deployment (R4). Specific mentioning of operation type (R3) and scenario downscaling (R1) are outside its scope. Dehbozorgi et al. \cite{Dehbozorgi2025} develop a \ac{SE}-based detection scheme for false data injection attacks. They introduce cost functions for misdetection and optimize alarm thresholds for each meter using an autoencoder and largest normalized residual test. This work shows how measurement placement and thresholding affect detection costs. It relies on an \ac{SE} (R4) and ensures anomalies are caught under limited data (R5). Idlbi and Graeber \cite{en19030720} propose a digital twin framework for \ac{LV} distribution grids with high \ac{DER} share, integrating \ac{SE} with congestion management in a unified operational environment. The digital twin continuously estimates the grid state from heterogeneous measurements and uses the estimated state to identify and resolve thermal and voltage limit violations. It directly uses \ac{SE} under realistic measurement conditions (R4) and congestion events are detected via estimated state (R5). Besides that, congestion management is an explicit operational objective (R3). The framework does not perform macro-level scenario downscaling (R1), and no standardized benchmark grids are used (R6). Koch et al.\ \cite{kochSmartMetersGrid2023} evaluate the substitution of topology-centred measurement infrastructure by smart meters for grid state identification in six SimBench \ac{LV} grids, showing that substitution becomes feasible at deployment rates between 30\% and 60\% when both voltage and current deviations are considered. They further integrate this approach into an agent-based local energy and flexibility market use case. Asman et al.\ \cite{asmanExpectedDeviationsGrid2025} extend this line of work by quantifying expected deviations in grid state identification using a \ac{SGIM} on the same six SimBench \ac{LV} grids. With SGIM measurements at all transformer distribution lines, voltage deviations range from 0.27\% to 0.58\% and current deviations from 1.69\% to 9.19\%. With only a single measurement at the transformer supply line, the deviations significantly exceed the 0.5\% voltage accuracy target in five of six grids, rendering the grid state identification unsuitable for curative congestion management. These findings address grid state identification with sparse measurements (R4) and current deviations indicate limits of congestion detectability (R5), but the analysis does not include scenario-based \ac{DER} projections (R1--R3) and uses SimBench grids (R6). Von der Heyden et al.\ \cite{vonderheydenPrivacyPreservingPowerFlow2025} address a complementary challenge by proposing a privacy-preserving power flow analysis via secure multi-party computation, enabling grid state identification without exposing individual prosumer data. This work tackles the data privacy constraints that limit measurement availability in practice (R4), but does not address congestion detection (R5) or scenario-based analysis (R1--R3, R6).

\subsection{Summary of the Analysis of Related Work and Research Gap}\label{subsec:ResearchGap}
The systematic review of the literature, summarized in \autoref{tab:overview_related_work}, reveals a significant fragmentation in current research regarding the monitoring of \ac{LV} grids, consistent with the broader literature gaps identified by \cite{wagnerSystematicReviewState2026}. While each requirement (R1--R6) is addressed individually or in small clusters by existing works, no single study provides an integrated approach that spans the entire chain from macro-level scenario downscaling to observability-aware congestion detection.

\begin{table*}[tp]
\centering
\caption{Summary of R1--R6 fulfillment for reviewed works across three key categories.\\
\small{\fullcirc~fulfilled, \halfcirc~partially fulfilled, \emptycirc~not fulfilled.}}
\label{tab:overview_related_work}
\small
\begin{tabularx}{\textwidth}{>{\raggedright\arraybackslash}X c c c c c c}
\toprule
\textbf{Author} & \textbf{R1} & \textbf{R2} & \textbf{R3} & \textbf{R4} & \textbf{R5} & \textbf{R6} \\ \midrule

\multicolumn{7}{l}{\textit{Scenario Downscaling and Benchmarking}} \\ \midrule
Luo et al. \cite{Luo2026} & \halfcirc & \emptycirc & \emptycirc & \emptycirc & \emptycirc & \emptycirc \\
Fakhrooeian et al. \cite{Fakhrooeian2024} & \fullcirc & \fullcirc & \emptycirc & \emptycirc & \emptycirc & \fullcirc \\
Treutlein et al. \cite{Treutlein2026} & \emptycirc & \emptycirc & \emptycirc & \emptycirc & \emptycirc & \fullcirc \\
Meinecke et al. \cite{Meinecke2020} & \emptycirc & \emptycirc & \emptycirc & \emptycirc & \emptycirc & \fullcirc \\ \midrule

\multicolumn{7}{l}{\textit{Grid Congestion Analysis and Operational Flexibility}} \\ \midrule
Damianakis et al. \cite{Damianakis2025} & \emptycirc & \fullcirc & \fullcirc & \emptycirc & \emptycirc & \emptycirc \\
Protopapadaki and Saelens \cite{Protopapadaki2017} & \emptycirc & \fullcirc & \emptycirc & \emptycirc & \emptycirc & \emptycirc \\
Delchambre et al. \cite{Delchambre2023} & \emptycirc & \fullcirc & \fullcirc & \emptycirc & \emptycirc & \emptycirc \\
Khan et al. \cite{Khan2026} & \emptycirc & \emptycirc & \fullcirc & \emptycirc & \emptycirc & \emptycirc \\ \midrule

\multicolumn{7}{l}{\textit{Observability and Detectability}} \\ \midrule
Fotopoulou et al. \cite{Fotopoulou2022} & \emptycirc & \emptycirc & \emptycirc & \fullcirc & \halfcirc & \emptycirc \\
Mattoo et al. \cite{Mattoo2025} & \emptycirc & \emptycirc & \emptycirc & \fullcirc & \emptycirc & \emptycirc \\
Paruta et al. \cite{Paruta2021} & \emptycirc & \emptycirc & \emptycirc & \fullcirc & \emptycirc & \emptycirc \\
Buason et al. \cite{Buason2024} & \emptycirc & \emptycirc & \emptycirc & \fullcirc & \fullcirc & \emptycirc \\
Dehbozorgi et al. \cite{Dehbozorgi2025} & \emptycirc & \emptycirc & \emptycirc & \fullcirc & \fullcirc & \emptycirc \\
Idlbi and Graeber \cite{en19030720} & \emptycirc & \halfcirc & \halfcirc & \fullcirc & \halfcirc & \emptycirc \\
Koch et al. \cite{kochSmartMetersGrid2023} & \emptycirc & \emptycirc & \halfcirc & \fullcirc & \halfcirc & \fullcirc \\
Asman et al. \cite{asmanExpectedDeviationsGrid2025} & \emptycirc & \emptycirc & \emptycirc & \fullcirc & \halfcirc & \fullcirc \\
Von der Heyden et al. \cite{vonderheydenPrivacyPreservingPowerFlow2025} & \emptycirc & \emptycirc & \emptycirc & \fullcirc & \emptycirc & \emptycirc \\ \bottomrule
\end{tabularx}

\end{table*}

The core research gap lies in the disconnect between long-term grid planning and operational \ac{SE}. While scenario-based studies often provide detailed pathways for \ac{DER} expansion, they typically assume a "perfectly observable" grid. Conversely, \ac{SE} research focuses on algorithmic accuracy but frequently lacks the context of evolving congestion patterns resulting from specific energy transition pathways. To bridge this gap, this work proposes a methodological framework that links the German Federal Government energy transition pathway for the period 2025–2045 with a systematic evaluation of \ac{SE} performance. This evaluation integrates scenario-based downscaling (R1) and physical grid simulation (R2) with an observability-aware \ac{SE} (R4). To establish a consistent reference framework for evaluating the detection limits of the \ac{SE}, active congestion-management measures (e.g., curtailment) are deliberately excluded from the simulation (R3). This "unmitigated baseline" approach ensures that the detectability of thermal and voltage limits (R5) is assessed against the full spectrum of potential grid violations. By utilizing reproducible benchmark grids (R6), this work provides \acp{DSO} with quantitative guidelines for configuring their monitoring infrastructure (the choice of transformer- and feeder-level instrumentation alongside smart-meter density) as \ac{DER} penetration grows, ensuring that reliable \ac{SE} remains feasible as distribution grids transition from passively operated to actively managed systems.

%% file: 3_mathematical_analysis_state_estimation.tex
\section{State Estimation: Metrics and Algorithm} \label{sec:mathematical_analysis_state_estimation}

\ac{LV} networks typically lack sufficient measurement infrastructure to achieve full observability. The number of installed sensors rarely meets the mathematical minimum for \ac{WLS} \ac{SE}, requiring pseudo-measurements derived from load profiles or consumption models to fill the gap \cite{Radhoush2022, Vijaychandra2023, Fotopoulou2022, wagnerSystematicReviewState2026}. The accuracy of the resulting state estimate depends on the number, location, and type of real measurements as well as the quality of the pseudo-measurement model \cite{Netto2022, Paruta2021, Buason2024}.

\subsection{Definition and Application of Suitable Quality Metrics}\label{subsec:definition_application_quality_metrics}

Defining suitable quality metrics for \ac{LV} distribution grid \ac{SE} is not trivial. In contrast to transmission system \ac{SE}, the objective in \ac{LV} grids is typically not a perfectly accurate reconstruction of the full system state, but rather the reliable detection of operationally critical conditions such as overloads and voltage limit violations.

A practical basis for the definition of accuracy metrics is provided by \ac{VDEFNN} guidance on \ac{LV} grid \ac{SE} \cite{FNN2021, VDEFNN2024StateEstimationLV}. In this context, accuracy targets are formulated explicitly for both voltage and current estimation, and evaluated using robust statistical measures. The voltage estimation accuracy $f_V$ is defined as the absolute deviation between the estimated node voltage $V_{\mathrm{est}}$ and the true voltage $V_{\mathrm{real}}$, normalized by the nominal voltage $V_n$:
\begin{equation}
    f_V = \left|\frac{V_{\mathrm{est}} - V_{\mathrm{real}}}{V_n}\right| \cdot 100\%.
\end{equation}
The study investigates a voltage accuracy threshold of $2\%$ (including measurement inaccuracies). For evaluation, the $99$th percentile of node voltage errors is used, which provides a robust metric that focuses on worst-case performance while remaining less sensitive to single extreme outliers.

Similarly, the current estimation accuracy $f_I$ is defined as the deviation between the estimated branch current $I_{\mathrm{est}}$ and the true current $I_{\mathrm{real}}$, normalized by the current-carrying capacity $I_z$:
\begin{equation}
    f_I = \left|\frac{I_{\mathrm{est}} - I_{\mathrm{real}}}{I_z}\right| \cdot 100\%.
\end{equation}
For current estimation, the study investigates a target accuracy of $10\%$ and evaluates the $99$th percentile of the current error. In addition, the evaluation is restricted to critical network segments that experience high loading under relevant operating scenarios, reflecting the practical goal of reliably detecting overload risks. The \ac{VDEFNN} targets therefore represent a practically relevant benchmark if the objective of \ac{SE} is the avoidance of critical grid states.

For tracking the trend of voltage estimation quality as a function of measurement penetration, independently of regulatory targets, a complementary diagnostic metric is used. The \emph{mean normalized voltage error} $\bar{\varepsilon}$ averages the pointwise voltage deviation across all $N_\text{LV}$ \ac{LV} nodes:
\begin{equation}\label{eq:norm_voltage_err}
    \bar{\varepsilon} \;=\; \frac{1}{N_\text{LV}} \sum_{i=1}^{N_\text{LV}}
    \left| \frac{\hat{v}_i - v_{\text{true},i}}{\max_j v_{\text{true},j}} \right| \cdot 100\%,
\end{equation}
where $\hat{v}_i$ is the \ac{SE} voltage estimate at node $i$ and $v_{\text{true},i}$ the simulation ground truth. Normalization by the snapshot maximum $\max_j v_{\text{true},j}$ rather than by the nominal voltage $V_n$ eliminates systematic slack offsets that arise when the substation reference voltage deviates from nominal, and renders the metric sensitive to the \emph{shape} of the estimated voltage profile rather than its absolute level. Unlike $f_{V,\text{p99}}$, which targets worst-case compliance, $\bar{\varepsilon}$ summarizes the network-wide average behavior and is therefore suited for sweep analyses in which the measurement penetration is varied continuously.

Beyond pure numerical deviation metrics, it is also useful to evaluate the detection performance of the estimator with respect to critical grid states. In this context, sensitivity (true positive rate) and specificity (true negative rate) provide meaningful indicators for assessing how reliably overloads or voltage violations are detected while avoiding false alarms.

In practice, reference values $V_{\mathrm{real}}$ and $I_{\mathrm{real}}$ are rarely available across the full network, limiting direct error computation to the few sensor locations. Alternative validation strategies include residual analysis \cite{Guo2013}, leave-one-out cross-validation \cite{LorenzMeyer2023}, and simulation-based evaluation against a known ground truth \cite{AbdelMajeed2015}. This work adopts the simulation-based approach: congestion scenarios from \autoref{sec:impacts_development_pathways_load_limits} provide the ground truth, and the \ac{SE} is evaluated against the full power-flow solution. Online uncertainty quantification via the gain matrix \cite{strobelUncertaintyQuantificationBranchCurrent2024} complements the offline evaluation.

\subsection{The BC-Mod Branch-Current WLS Estimator}\label{subsec:bc_mod_estimator}

The \ac{SE} implementation evaluated in \autoref{sec:state_estimation_quality} is based on the modified branch-current \ac{WLS} estimator \ac{BC-Mod} \cite{strobelUncertaintyQuantificationBranchCurrent2024, ipachModifiedBranchCurrentBased2021}. In contrast to the node-voltage \ac{WLS} formulation, \ac{BC-Mod} uses complex branch currents as the state variables. This results in a linear measurement Jacobian whose entries depend only on the grid parameters, not on the current measurement set. Compared to node-voltage \ac{SE}, \ac{BC-Mod} is therefore less sensitive to measurement errors and pseudo-measurement inaccuracies, a property that is particularly advantageous for \ac{LV} grids with high pseudo-measurement shares.

The \ac{WLS} state vector $\hat{\mathbf{x}}$ is constructed from the real and imaginary parts of the complex branch currents for all $M$ lines across all three phases. The estimator minimizes the weighted sum of squared measurement residuals:
\begin{equation}
    J(\hat{\mathbf{x}}) = \left(\mathbf{z} - \mathbf{H}\hat{\mathbf{x}}\right)^\top \mathbf{W} \left(\mathbf{z} - \mathbf{H}\hat{\mathbf{x}}\right),
\end{equation}
where $\mathbf{z}$ is the measurement vector, $\mathbf{H}$ is the linear measurement Jacobian, and $\mathbf{W} = \mathbf{R}^{-1}$ is the weight matrix derived from the measurement covariance matrix $\mathbf{R}$. Measurements (substation phasor measurements, smart meter readings) and pseudo-measurements (load profile estimates) enter $\mathbf{z}$ with weights reflecting their respective uncertainties. In this work, real measurements are assigned $\sigma_\text{real} \approx 1.7\%$ and pseudo-measurements $\sigma_\text{pseudo} = 40\%$ (see \autoref{subsec:data_smgw_penetration_se}).

Node voltages are not part of the state vector but are recovered in a subsequent forward sweep. Starting from the substation voltage, the voltage drop over each branch is calculated as
\begin{equation}
    \mathbf{v}_{\mathrm{drop},l,k} = \mathbf{Z}_{l,k} \cdot \mathbf{i}_{l,k},
    \qquad
    \mathbf{v}_l = \mathbf{v}_k - \mathbf{v}_{\mathrm{drop},l,k},
\end{equation}
where $\mathbf{Z}_{l,k}$ is the three-phase impedance matrix of branch $l$--$k$. The distinguishing feature of \ac{BC-Mod} over the conventional branch-current \ac{SE} \cite{ipachModifiedBranchCurrentBased2021} is a backward sweep that incorporates available node voltage measurements: each measured voltage is back-propagated to infer a substation voltage estimate, and the average of these estimates initializes the next forward sweep. This iterative process achieves accuracy comparable to nonlinear \ac{SE} formulations at significantly lower computational cost.

The implementation also provides \ac{UQ} for the estimated branch currents and node voltages \cite{strobelUncertaintyQuantificationBranchCurrent2024}. Current uncertainty is derived from the diagonal of the gain matrix $\mathbf{G} = \left(\mathbf{H}^\top \mathbf{W} \mathbf{H}\right)^{-1}$, which equivalently serves as the covariance matrix of the \ac{WLS} state vector. Voltage uncertainty is propagated analytically through the forward-sweep equations using the branch impedances and the current covariances. Both uncertainties are expressed as 95\% credibility intervals, supporting manual decision-making in non-routine grid operation \cite{wagnerSupportingManualDecisionMaking2025}.

%% file: 4_development_pathways_load_limits.tex
\section{Impacts of the Energy Transition on Load Limits in Low-Voltage Grids} \label{sec:impacts_development_pathways_load_limits}

This section quantifies congestion in low-voltage grids under the Federal Government energy transition pathway (2025--2045), covering scenario definition (\autoref{subsec:data_basis}), simulation input data (\autoref{subsec:Simulation_Input_Data}), methodology (\autoref{subsec:concept_modeling}), and evaluation (\autoref{subsec:evaluation_part_1}).

\subsection{Energy Transition Pathway}\label{subsec:data_basis}

All simulations are based on the German Federal Government deployment trajectory for four technology categories: (i) publicly accessible charging stations, (ii) \acp{EV}, (iii) heat pumps, and (iv) rooftop \ac{PV} capacity. The 2025 values are the measured baseline and the 2035 and 2045 values the official targets, with the intermediate years 2030 and 2040 obtained by linear interpolation. Comparable studies by Agora Energiewende \cite{agora2024} and the Fraunhofer Institute for Solar Energy Systems \cite{fraunhoferISE2024} arrive at broadly similar deployment trajectories. The Federal Government figures are drawn from the Renewable Energy Sources Act \cite{eeg2023}, the Photovoltaic Strategy \cite{bmwk2023photovoltaik}, the Master Plan for Charging Infrastructure II \cite{bmdv2022ladeinfrastruktur}, and the Heat Pump Offensive report \cite{bmdv2022waermepumpen}. Per-100-\ac{HH} indicators are calculated using \cite{bbsranalysen2024}, which assumes 42.0 million households in 2025 and 42.6 million in 2045. The resulting scenario values are summarised in \autoref{tab:energy-transition-pathways}.

\begin{table*}[tp]
\centering
\begin{threeparttable}
\caption{Federal Government national deployment trajectory for public charging stations, \acp{EV}, rooftop solar \ac{PV}, and heat pumps for the years 2025, 2030, 2035, 2040, and 2045 (units per row). The 2025 column is the measured baseline. Each Source entry lists the baseline reference (2025) followed by the target reference (2035 and 2045). Values for 2030 and 2040 (shown in italics) are obtained by linear interpolation between the 2025, 2035, and 2045 anchor values. The trajectory sets the temporal growth. The spatial allocation per network is settlement-specific (see \autoref{subsec:concept_modeling}).}
\label{tab:energy-transition-pathways}
\footnotesize
\renewcommand{\arraystretch}{1.3}
\begin{tabularx}{\textwidth}{@{} l l l *{5}{>{\centering\arraybackslash}X} @{}}
\noalign{\hrule height 1.1pt}
\textbf{Category} & \textbf{Source} & \textbf{Unit} & \textbf{2025} & \textbf{2030} & \textbf{2035} & \textbf{2040} & \textbf{2045} \\
\noalign{\hrule height 0.8pt}
Public Charging Stations & \cite{bundesnetzagentur2026, nationaleleitstelle2024} & \#/100 HH & 0.22 & \textit{1.10} & 1.99 & \textit{2.85} & 3.72 \\
Electric Vehicles       & \cite{kraftfahrtbundesamt2025, anzahlBEV2024}          & \#/100 HH & 3.93 & \textit{28.0} & 52.0 & \textit{68.0} & 84.1 \\
Rooftop Solar \ac{PV}  & \cite{bundesnetzagentur2026anzahlpv, eeg2023}                & GW        & 76.0 & \textit{115.5} & 155 & \textit{200.5} & 246 \\
Heat Pumps              & \cite{bundesverbandwaermepumpe2026, systementwicklungsstrategie2024} & \#/100 HH & 3.81 & \textit{13.9} & 24.0 & \textit{33.0} & 42.0 \\
\noalign{\hrule height 1.1pt}
\end{tabularx}
\end{threeparttable}
\end{table*}

\subsubsection{Baseline (2025)}\label{subsubsec:baseline}
According to \cite{bundesnetzagentur2026}, Germany had 91,351 publicly accessible charging stations (comprising 165,295 individual charging points) as of 1 January 2025, corresponding to approximately 0.22 stations per 100~HH. In \cite{kraftfahrtbundesamt2025}, 1,651,643 \acp{EV} are reported, equivalent to 3.93 per 100~HH. After wind energy, solar \ac{PV} is Germany's second largest renewable source. In 2025, 4,474,362 rooftop systems with a combined capacity of 76.0~GW were in operation \cite{bundesnetzagentur2026anzahlpv}, corresponding to 10.7 systems per 100~HH. At the end of 2024, approximately 1,600,000 heat pumps were installed \cite{bundesverbandwaermepumpe2026}, equivalent to 3.81 per 100~HH.

\subsubsection{Public Charging Stations}\label{subsubsec:public_charging_points}
The Federal Government targets 1.38--2.60 charging stations per 100~HH by 2035 and approximately 3.72 by 2045 \cite{nationaleleitstelle2024}, complemented by the build-out of a high-power charging network and grid-coordinated charging. Private charging infrastructure is a complementary buffer that limits peak demand on the public grid \cite{nationaleleitstelle2024}.

\subsubsection{Electric Vehicles}\label{subsubsec:electric_vehicles}
The Federal Government targets approximately 21.9 million \acp{EV} (52.0 per 100~HH) by 2035 and 35.8 million (84.1 per 100~HH) by 2045 \cite{anzahlBEV2024}. These figures imply a large-scale electrification of the passenger car fleet, reaching roughly 70\% of the current stock by 2045. Key instruments include direct subsidies, the anticipated 2035 zero-emission mandate for new registrations, and rising CO\textsubscript{2} pricing.

\subsubsection{Rooftop Solar PV}\label{subsubsec:rooftop_solar_pv}
Federal Government targets for rooftop \ac{PV} grow from 76.0~GW in 2025 to 155~GW in 2035 and approximately 246~GW in 2045 \cite{eeg2023}. Capacity roughly doubles from 2025 to 2035 and grows by a further 60\% to 2045. These values are the rooftop share of the broader Federal PV expansion, which also includes ground-mounted systems.

\subsubsection{Heat Pumps}\label{subsubsec:heat_pumps}
Heat pump deployment rises from 3.81 per 100~HH in 2025 to 24.0 in 2035 and 42.0 in 2045, reflecting a major shift toward electrified individual space heating \cite{systementwicklungsstrategie2024}. Key drivers are rising CO\textsubscript{2} prices under the EU ETS and anticipated reductions in network charges and electricity taxes \cite{agora2024}.

\subsection{Simulation Input Data}\label{subsec:Simulation_Input_Data}
Building on the values in \autoref{tab:energy-transition-pathways}, intermediate years were completed via linear interpolation and the 2025 column is anchored to the empirically observed baseline. Subsequently, existing load profiles were compiled for public \cite{en16062619} and private charging infrastructure \cite{privatelastprofile}, rooftop solar \ac{PV} generation \cite{PFENNINGER20161251,STAFFELL20161224}, households \cite{haushaltslastprofil}, and heat pumps \cite{waermepumpelastprofil}. These load profiles were harmonized to represent a full non-leap year (365 days) at a 15‑minute temporal resolution. This results in 35,040 intervals, each associated with an average load in kW for the respective 15‑minute period.

For public charging stations, in addition to standard weekday and weekend profiles, outliers observed in the raw data were explicitly modeled. The modeling utilizes load profiles representing the 70th to 85th percentiles to approximate typical demand conditions. To capture potential peak-load events, a specific high-power profile based on maximum observed utilization was integrated for individual charging units across all scenarios. Distinct load profiles for rural and urban areas were available \cite{en16062619}. For private charging infrastructure, the underlying load profile assumed a daily plug-in factor of 70\% with a charging demand of 3.68 kW (16 A, 230 V). Based on empirical measurements, this implies that only 70\% of all \acp{EV} are charged on a given day \cite{privatelastprofile}. For household demand, the newly updated H25 standard load profile was used. This profile reflects modern consumption patterns by incorporating structural shifts in energy use over the past 25 years to provide a more precise baseline for current residual demand. To adjust the reference single-family household profile to seasonal variation, a dynamization function was applied \cite{haushaltslastprofil}. The original normalization to an annual consumption of 1,000,000 kWh was rescaled to a baseline of approximately 3,500 kWh per household using a linear adjustment. This value is above the average annual electricity consumption for German households, for a conservative estimate of residential demand \cite{EnergyConsumptionHousehold}. The authors in \cite{WohnungenimMFH} report that a multi-family building in Germany comprises, on average, 6.76 apartments. This value was rounded down to 6, and the single-family household load profile was scaled by this factor to approximate a multi-family building load profile.

For the \ac{PV} generation profile, Germany’s average global horizontal irradiation of 1,086 kWh/m² \cite{GlobalstrahlungDeutschland} motivated the selection of Frankfurt am Main as the reference location. Frankfurt exhibits approximately 1,100 kWh/m², closely matching the national average \cite{GlobalstrahlungFrankfurt}. A typical \ac{PV} system size of 11.5 kWp was assumed \cite{DurchschnittlichePV}. Based on data from \cite{AusrichtungPV}, the following distribution of \ac{PV} array orientations was adopted: south (62.3\%), west (21.0\%), and east (16.7\%), obtained by excluding north-facing systems and scaling the remaining south, west, and east shares to 100\% of the simulated capacity. To account for regional differences in building stock, the electric load of urban heat pumps was assumed to be four times that of rural heat pumps. This factor of four accounts for the difference in heating loads between urban multi-family buildings and rural single-family houses \cite{iwu2015}.

\subsection{Methodology}
\label{subsec:concept_modeling}
This section investigates when and how severely grid congestions occur in \ac{LV} distribution networks under the Federal Government energy transition pathway, considering three levels of grid equipment quality and passive network operation. The study applies time-series power-flow analysis based on standardized SimBench reference networks, scenario-specific equipment at \acp{NCP}, and a formalized detection and classification of congestion events.

Active congestion-management measures (storage systems, bidirectional charging, load shifting, dynamic tariffs, intelligent control strategies) are deliberately excluded. At each NCP, PV generation and demand are represented as separate elements at the same bus and netted by the power flow, so local generation offsets local demand and only the residual is exchanged with the grid. The analysis thus quantifies when purely passive operation reaches its limits.

\subsubsection{Selection and Modeling of Network Structures}
\begin{figure*}[tp]
  \centering
  \includegraphics[width=0.9\linewidth]{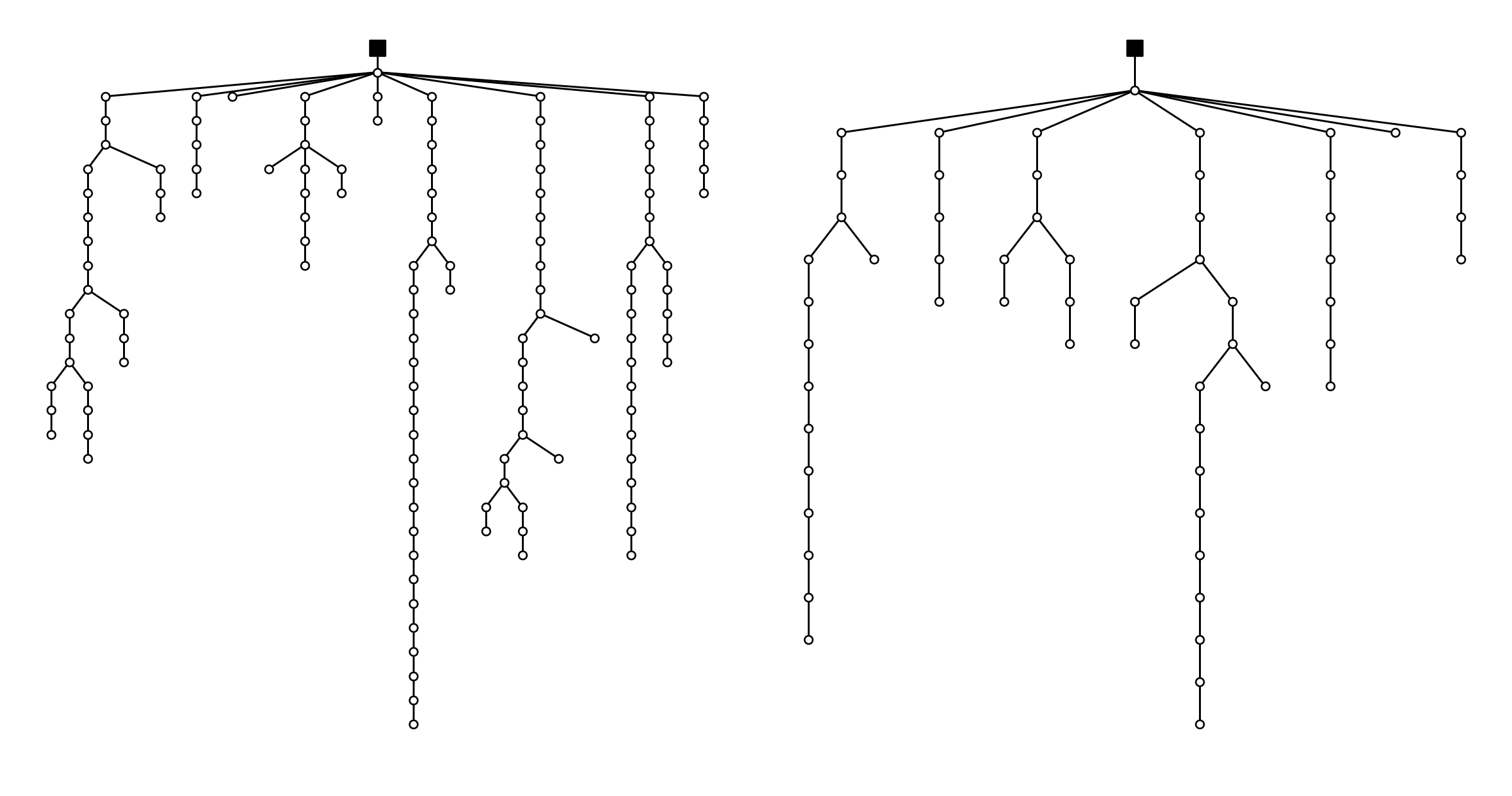}
  \caption{Schematic visualization of the two LV SimBench networks used in this work: \texttt{1-LV-rural3--0-no\_sw} (left) and \texttt{1-LV-urban6--0-no\_sw} (right).}
  \label{fig:network_visualization}
\end{figure*}

The study is based on standardized \ac{LV} reference networks from the SimBench dataset \cite{SimBench2020}. To represent contrasting settlement structures, two different networks are used. For the simulation of urban scenarios, the open-access SimBench code \texttt{1-LV-urban6--0-no\_sw} is used. It is characterized by densely built areas with predominantly multi-family homes. The rural scenarios are based on the SimBench code \texttt{1-LV-rural3--0-no\_sw}, which represents single-family home areas with low connection density. Both networks are shown in \autoref{fig:network_visualization}.

The network topology (lines, transformers, buses) and all electrical parameters of the original SimBench data correspond to the \emph{good} equipment level described below and remain unchanged for that level, apart from a minimum-impedance floor. All load and generation elements originally contained in the reference networks are removed and fully replaced by scenario-based load and generation data to ensure that observed congestions result exclusively from the defined scenarios.\footnote{A minimum line resistance and reactance of $0.001~\Omega$ is enforced to avoid quasi-zero impedances in the SimBench data. These values are orders of magnitude below realistic line impedances and do not materially affect results.}

To investigate the sensitivity of congestion patterns to the existing network infrastructure, three equipment quality levels are considered. These levels vary the distribution transformer rating and the cable type (and thus the thermal current limit) while preserving the network topology, i.e., all bus positions, line lengths, and connection points remain identical. \autoref{tab:equipment-levels} summarizes the parameterization for both area types.

\begin{table}[htbp]
\centering
\caption{Grid equipment quality levels for rural and urban low-voltage networks.}
\label{tab:equipment-levels}
\footnotesize
\begin{tabular}{l l c l c}
\hline
\textbf{Level} & \textbf{Area} & \textbf{Transformer} & \textbf{Cable type} & \textbf{Thermal limit} \\
\hline
Good    & Rural & 400\,kVA & NAYY 4$\times$150 & 270\,A \\
Good    & Urban & 630\,kVA & NAYY 4$\times$240 & 357\,A \\
\hline
Medium  & Rural & 250\,kVA & NAYY 4$\times$120 & 242\,A \\
Medium  & Urban & 400\,kVA & NAYY 4$\times$150 & 270\,A \\
\hline
Poor    & Rural & 160\,kVA & NAYY 4$\times$50  & 142\,A \\
Poor    & Urban & 250\,kVA & NAYY 4$\times$120 & 242\,A \\
\hline
\end{tabular}
\end{table}

The \emph{good} level corresponds to the original SimBench default parameterization and thus to the transformer ratings and cable cross-sections specified in the respective reference networks. The \emph{medium} and \emph{poor} levels represent progressively weaker infrastructure by reducing both the transformer rating and the cable cross-section. The selected transformer ratings are consistent with real German \ac{LV} grids. Across 87 analyzed real-world networks, characteristic transformer size distributions can be observed for both rural and urban grids \cite{KerberWitzmann2008}. The ratings used in this work therefore correspond to the three most frequent types for the respective settlement structures. The chosen cable cross-sections likewise reflect differentiated rural and urban equipment levels, with 150, 120, and 50~mm\textsuperscript{2} (good, medium, and poor) for rural networks and 240, 150, and 120~mm\textsuperscript{2} for urban networks \cite{LindnerEtAl2016,WintzekEtAl2021}. In the \emph{poor} rural case, for example, the transformer capacity is reduced by 60\,\% relative to the \emph{good} baseline, and the cable thermal limit decreases by approximately 47\,\%. These variations allow the analysis to capture the range of grid headroom that may exist in practice, as real \ac{LV} grids exhibit considerable heterogeneity in their installed equipment.

\subsubsection{Scenario Setup}
Future developments are represented by the Federal Government energy transition pathway described in \autoref{subsec:data_basis}. The current state in 2025 serves as the reference. The target years are 2030, 2035, 2040, and 2045. Each year is analysed for both the urban and the rural network, yielding the ten scenarios listed in \autoref{tab:scenarios}.

\begin{table}[htbp]
\centering
\caption{Simulated test scenarios by area type and year.}
\label{tab:scenarios}
\begin{tabular}{r l r}
\hline
Scenario & Area type & Year \\
\hline
1 & rural & 2025 (baseline) \\
2 & rural & 2030 \\
3 & rural & 2035 \\
4 & rural & 2040 \\
5 & rural & 2045 \\
6 & urban & 2025 (baseline) \\
7 & urban & 2030 \\
8 & urban & 2035 \\
9 & urban & 2040 \\
10 & urban & 2045 \\
\hline
\end{tabular}
\end{table}

Each of the ten scenarios is simulated for all three equipment quality levels defined in \autoref{tab:equipment-levels}, yielding a total of \mbox{$10 \times 3 = 30$} scenario-equipment combinations.

\subsubsection{Mapping of Network Connection Points and Component Modeling}
\acp{NCP} are uniquely mapped to the load buses of the respective SimBench network. This preserves the spatial distribution of loads while ensuring a consistent one-to-one assignment between \acp{NCP} and buses. If a direct assignment is not possible, \acp{NCP} are assigned to available free load buses as a documented fallback. \acp{NCP} that cannot be assigned are not simulated.

In the rural network with 109 \acp{NCP}, each \ac{NCP} corresponds to one household (single-family home). In the urban network, one \ac{NCP} is an aggregate of multiple dwelling units (multi-family home). It includes 53 \acp{NCP}. The SimBench topologies provide 118 (rural) and 53 (urban) load buses. In the rural network, 109 of these are assigned to NCPs and the remaining 9 stay load-free throughout all simulated time steps, while in the urban network all 53 load buses are assigned.

Depending on the scenario, the following elements are modeled per single-family or multi-family home and assigned based on input data sheets including household electricity demand, public charging stations, \acp{EV}, rooftop solar \ac{PV} generation, and heat pumps. For clarity, public charging stations are coupled directly to households rather than introducing additional intermediate buses.

\autoref{tab:energy-transition-pathways} defines the national deployment trajectory, which sets the temporal growth of each technology over the period 2025--2045. The allocation within each network is settlement-specific: starting from this trajectory, devices are distributed across the \acp{NCP} by randomized assignment up to the maximum per-\ac{NCP} configurations shown in \autoref{fig:Ausstattung_NCP}. A rural single-family home is assigned at least household demand and at most two \acp{EV}, one \ac{PV} system, one heat pump, and one charging station. An urban \ac{NCP}, which aggregates six dwellings, is assigned at most six \acp{EV}, one \ac{PV} system, one heat pump, and two charging stations. Intermediate configurations follow from the randomized allocation. This settlement-specific intensity reflects the higher per-home car ownership in single-family areas, about 1.3 cars per rural household against 0.8 in metropolitan areas \cite{MiD2017}, and the limited rooftop area per dwelling in multi-family buildings. Heat pumps are allocated per \ac{NCP}, as one heat pump serves one building. The two reference networks are therefore deliberately contrasting cases of settlement structure rather than a statistically averaged national sample.

\begin{figure}[ht!]
  \centering
  \resizebox{0.7\linewidth}{!}{\input{figures/Ausstattung_NCP_tikz}}
  \caption{Minimum and maximum per-\ac{NCP} equipment configurations for the urban and rural scenarios. An urban \ac{NCP} aggregates six dwellings. Its maximum configuration adds six \acp{EV}, one \ac{PV} system, one heat pump, and two charging stations. A rural \ac{NCP} is a single-family home. Its maximum configuration adds two \acp{EV}, one \ac{PV} system, one heat pump, and one charging station. The legend (bottom) identifies the \ac{EV}, heat pump, household, charging station, and \ac{PV} symbols.}
  \label{fig:Ausstattung_NCP}
\end{figure}

To account for voltage sensitivity of real-world devices, voltage-dependent \acrshort{ZIP} (constant impedance, current and power) load models with specified power factors are used. Household loads are represented as a mixture of constant-impedance, constant-current and constant-power components (ZIP = 20\% / 10\% / 70\%) with a power factor of $\cos\varphi=0.98$. Private \ac{EV} charging is modeled predominantly as near-constant power (ZIP = 5\% / 5\% / 90\%, $\cos\varphi=0.99$), consistent with the power-electronics-controlled behavior of typical chargers. Heat pump loads are modeled with ZIP = 10\% / 10\% / 80\% and $\cos\varphi=0.97$. Their reactive power demand is slightly higher than that of household loads, but constant-power behavior still dominates. \ac{PV} generation is modeled as active-power injection with $\cos\varphi=1.0$. Reactive power provision by inverters (e.g.\ $Q(U)$ or $\cos\varphi(P)$ control) is intentionally not considered in line with the passive business-as-usual assumption.

\subsubsection{Congestion Events}\label{subsubsec:congestion_events}
A time step is classified as a congestion period if at least one of the following conditions is met:
(i) thermal overload, i.e., loading of a line or transformer exceeding 100\%,
(ii) voltage-band violation with $U < 0.95$~p.u.\ or $U > 1.05$~p.u., or
(iii) non-convergence of the power-flow calculation.
For each congestion period, affected elements (lines, transformers and buses) are recorded together with the corresponding loading or voltage values.

To separately report near-limit operating states, a \emph{grey zone} is defined. The \emph{grey zone} includes line and transformer loadings between 100\% and 110\% and voltage deviations in the ranges 0.90--0.95~p.u.\ (undervoltage) and 1.05--1.10~p.u.\ (overvoltage). Operating states outside these ranges are classified as \emph{hard} congestions, i.e., loadings above 110\% or voltages $V < 0.90$~p.u.\ or $V > 1.10$~p.u.

The root-cause classification for congestions is performed per time step. If \ac{PV} generation exceeds total demand in the generation--load balance, the period is classified as generation-dominant. Otherwise it is classified as load-dominant. Thermal overloads are interpreted according to this dominance. Overloads under generation dominance are classified as generation-side congestions, whereas overloads under load dominance are classified as load-side congestions. Voltage violations follow the same balance-based label, which in practice coincides with the deviation direction: overvoltage ($V_{\max} > 1.05$~p.u.) arises under generation dominance, and undervoltage ($V_{\min} < 0.95$~p.u.) under load dominance. Non-convergent periods are handled separately and reported as \emph{not converged}, as affected elements cannot be uniquely identified in these cases.

\subsubsection{Simulation Setup}

For each of the 30 scenario-equipment combinations (10 scenarios $\times$ 3 equipment quality levels), a time-series simulation is performed over a full non-leap year with 15-minute resolution (35{,}040 time steps). At each time step all network elements are updated and a power-flow calculation is executed using pandapower (Newton--Raphson, maximum 50 iterations). Non-convergent time steps are classified as \emph{not converged} congestion events. No element-level overload data are available for these cases.

Detailed element outputs are retained only for time steps containing at least one congestion event. Key scenario-level metrics, including congestion period counts, loading extremes, temporal distributions, and element-wise overload shares, are aggregated across all retained time steps.

\subsubsection{Simulation Verification}
A targeted validation study was conducted using one urban and one rural test network with the same topology and number of \acp{NCP} as the main simulation. Synthetic input data for eight time steps were constructed with known demand and generation values, covering the full range of operating conditions: no-congestion baselines, isolated high loads from each technology (\ac{PV}, \acp{EV}, public charging, heat pumps), simultaneous extreme demand triggering solver non-convergence, and balanced high-load/high-generation cases. All eight test cases produced results consistent with the expected outcomes. These results confirm the correctness of scenario-to-network mapping, device penetration scaling, congestion detection, and root-cause classification.

\subsection{Grid-Side Congestion Under Passive Network Operation}\label{subsec:evaluation_part_1}

This section analyzes grid-side congestion in urban and rural LV networks under passive operation up to 2045, systematically varying the grid equipment level (\autoref{tab:equipment-levels}). Device penetration rates are summarized in \autoref{tab:energy-transition-pathways} (\autoref{subsec:data_basis}).

\subsubsection{Development of Congestion Frequency} \label{subsec:devlopmentcong}

\begin{figure*}[tp]
  \centering
  \includegraphics[width=\linewidth]{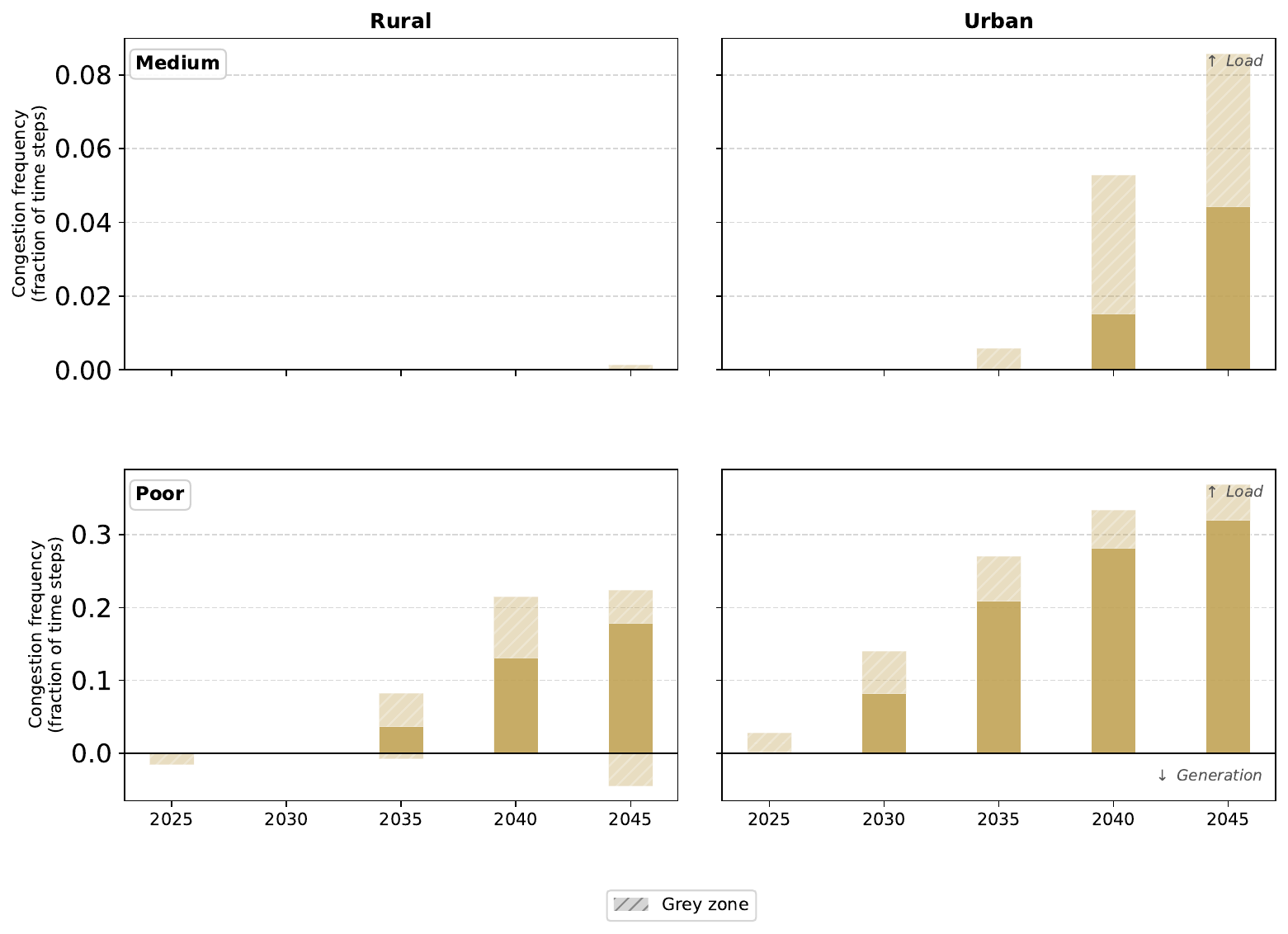}
  \caption{Development of congestion frequency in urban and rural grids by year, energy transition pathway, and equipment level.
           Bars show the fraction of all 35\,040 time steps per year with at least one congestion event,
           split into hard congestion (solid, loading $>$110\,\% or voltage $<$0.90 or $>$1.10\,\text{p.u.}) and
           grey-zone periods (hatched, 100--110\,\% or 0.90--0.95 or 1.05--1.10\,\text{p.u.}).
           Bars above the zero line indicate load-side, below it generation-side congestion.}
  \label{fig:Entwicklung_Engpasshaeufigkeit}
\end{figure*}

\autoref{fig:Entwicklung_Engpasshaeufigkeit} illustrates the development of the share of congestion periods in urban and rural grids over the period 2025--2045, differentiated by equipment level. A distinction is made between grey-zone periods and hard congestion events as defined in \autoref{subsubsec:congestion_events}.

Under good equipment, no congestion occurs throughout the entire simulation horizon in either network type. Under medium equipment, congestion first appears in 2035 in the urban network and in 2045 in the rural network. All observed events are exclusively load-driven. Under poor equipment, congestion is present from the 2025 baseline in both area types and intensifies substantially toward 2045, with hard congestion events increasingly dominating over grey-zone periods. In the rural network under poor equipment, generation-side congestion driven by PV feed-in first appears in 2025, whereas the urban network exhibits exclusively load-side congestion across all years. In the poor rural network, total congestion grows from 8.9\% of time steps in 2035 to 26.8\% by 2045 and is overwhelmingly load-driven, with the superimposed generation-side overvoltage signal (the downward bars) never exceeding about 4.5\% of time steps. This signal is threshold-sensitive: the peak midday bus voltage sits just at the 1.05 p.u. limit, dropping below it in 2030 (1.042 p.u.) and rising clearly above only in 2045 (1.066 p.u.). The 2030 valley is therefore congestion-free, with both regimes sub-threshold that year (peak transformer loading 95.4\%, minimum voltage 0.958 p.u.). In other words, 2030 falls in the window between the early generation-side overvoltage, a marginal grey-zone signal that recedes just below its threshold that year, and the onset of load-side congestion, which becomes dominant only from 2035 onward.

\subsubsection{Limit Violations at Network Components}

\begin{figure*}[tp]
  \centering
  \includegraphics[width=\linewidth]{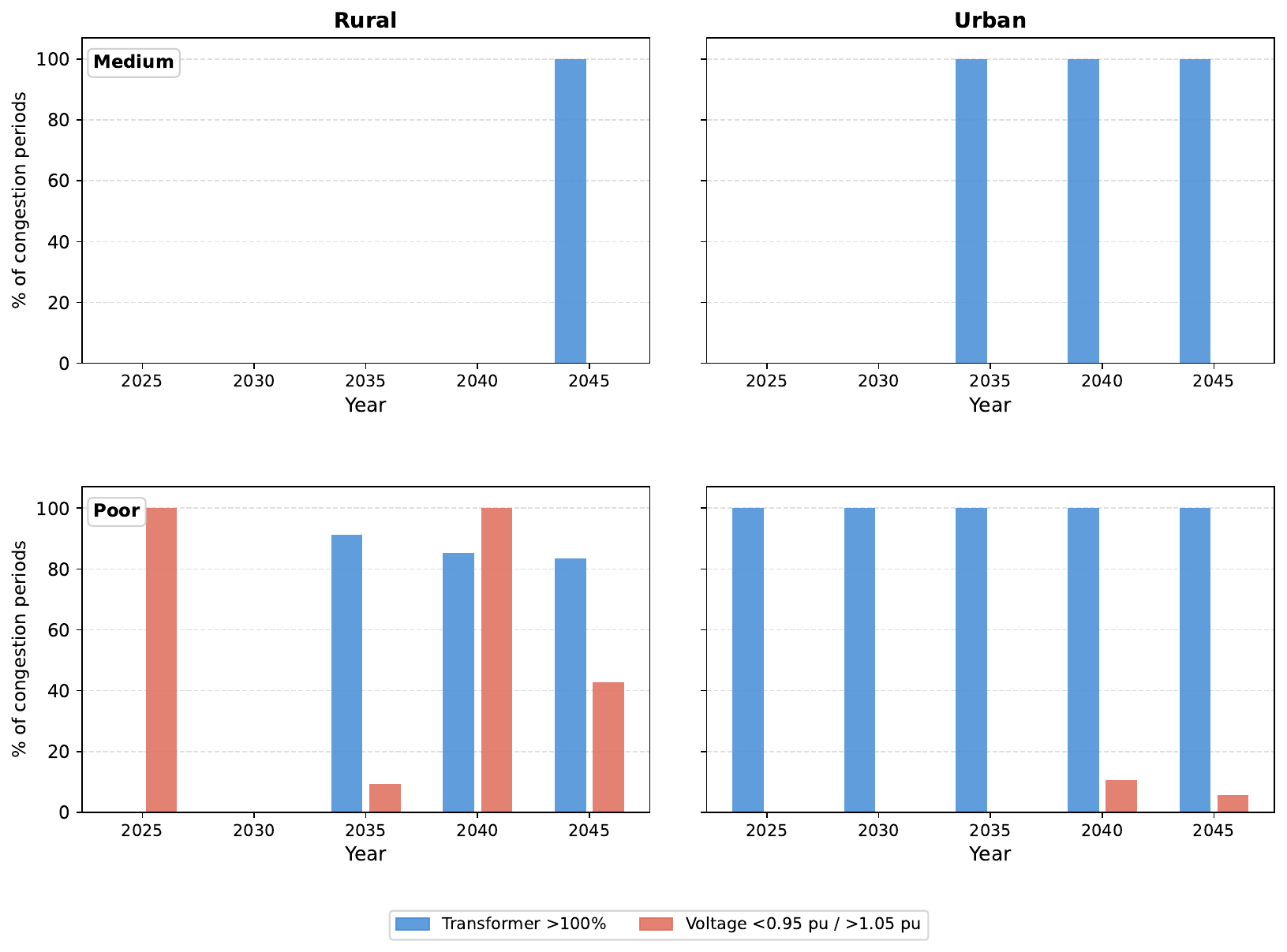}
  \caption{Component-level limit violations during congestion periods, by equipment quality level (rows: medium, poor) and network type (columns: rural, urban). For each year, two grouped bars show the fraction of congestion periods with transformer loading $>$100\,\% (blue) and the fraction with at least one bus voltage $<$0.95\,p.u.\ or $>$1.05\,p.u.\ (red). Empty bars indicate years without congestion events at that equipment level.}
  \label{fig:Grenzwertueberschreitungen}
\end{figure*}

\autoref{fig:Grenzwertueberschreitungen} quantifies which violation type drives congestion, showing for each year the fraction of congestion periods affected by transformer overload and by voltage-band violation.

\autoref{fig:Grenzwertueberschreitungen} confirms that congestion is driven by two distinct mechanisms: transformer overloading and voltage-band violations. Under medium equipment, transformer overloading is the sole violation type in urban networks from 2035, while rural congestion at 2045 is also driven exclusively by transformer overloading. Under poor equipment, transformer overloading dominates in urban networks across all years, consistent with the higher per-transformer demand density in multi-family-home areas. In rural networks, voltage-band violations are the primary mechanism in 2025 (PV-driven overvoltage), as the reduced transformer and cable capacity lowers the threshold at which PV backfeed causes voltage rise at peripheral nodes. From 2035 onward, rising \ac{EV} and heat pump penetration makes transformer overloading the most frequent violation type (present in 83 to 91\,\% of congestion periods), while the accompanying voltage violations shift in character from the PV-driven overvoltage of 2025 to load-side undervoltage at remote feeder ends. This undervoltage is not a secondary effect: it accompanies nearly every congestion period in 2040. By 2045 it affects a smaller share, as PV-driven overvoltage also returns at midday when the peak bus voltage again exceeds the 1.05 p.u. limit (see \autoref{subsubsec:seasonal_distribution}). No individual line exceeds its thermal rating in any of the evaluated scenarios (maximum observed: \mbox{$98.6\,\%$} in the 2040 urban network under poor equipment). The radial topology concentrates thermal stress at the transformer rather than distributing it across line segments.
\subsubsection{Congestion Severity}

\begin{figure*}[tp]
  \centering
  \includegraphics[width=\linewidth]{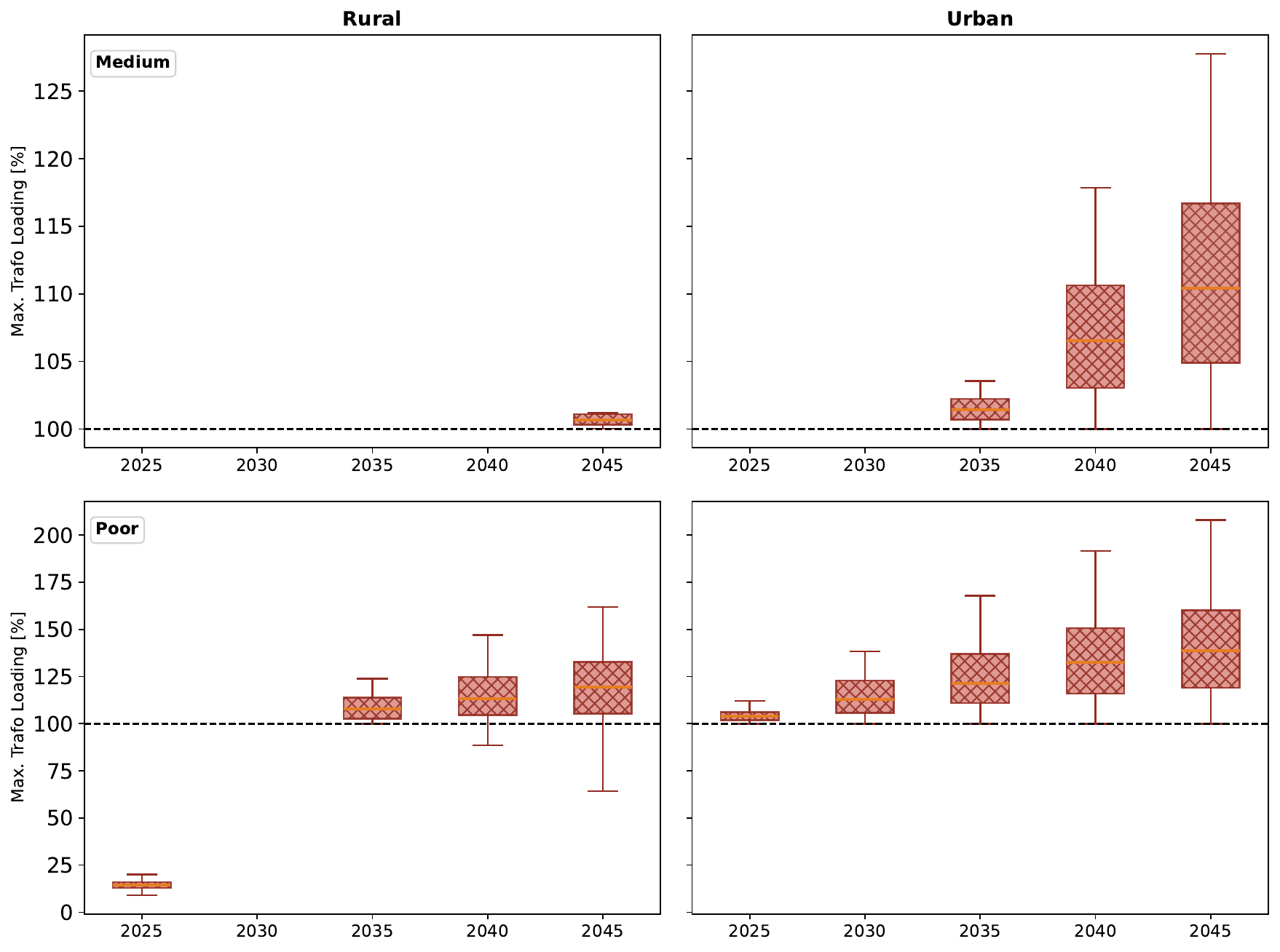}
  \caption{Distribution of maximum transformer loading during congestion periods, by year, equipment level, and area type.
           Only time steps with at least one congestion event are included.
           Box plots show the distribution per year, with separate panels for medium and poor equipment
           and both network types (rural/urban). Good equipment is omitted (no congestion events).
           Dashed line: 100\,\% overload threshold.
           Missing boxes indicate years without congestion events at that equipment level.}
  \label{fig:Engpassstaerke_Verteilung}
\end{figure*}

\autoref{fig:Engpassstaerke_Verteilung} extends the frequency analysis to severity, showing the distribution of maximum transformer loading exclusively during time steps classified as congestion events (\autoref{subsubsec:congestion_events}). While \autoref{fig:Entwicklung_Engpasshaeufigkeit} quantifies how often congestion occurs, \autoref{fig:Engpassstaerke_Verteilung} characterizes how severe it is. Good equipment is omitted from the figure because no congestion events occur under this equipment level in any of the ten scenarios, as established in \autoref{subsec:devlopmentcong}.

Under poor equipment, the distribution of transformer loading during congestion periods in \autoref{fig:Engpassstaerke_Verteilung} shifts markedly upward over time. In the urban network, the median loading rises from approximately 104\,\% in 2025 to approximately 139\,\% by 2045, with the interquartile range widening substantially, so congestion becomes both more frequent and more severe. The rural network follows the same trend from a lower base: its 2025 events are generation-side at low transformer loading, load-driven severity emerges only from 2035 and reaches about 119\,\% by 2045, and 2030 is congestion-free. Under medium equipment, congestion severity remains moderate, with the median only just above the 100\,\% threshold (about 110\,\% in the urban network by 2045).

\subsubsection{Seasonal Distribution of Congestion Periods}
\label{subsubsec:seasonal_distribution}

This section examines the seasonal structure of congestion events to identify in which months and under which conditions the load- and generation-side mechanisms dominate.

\begin{figure*}[tp]
  \centering
  \includegraphics[width=0.95\linewidth]{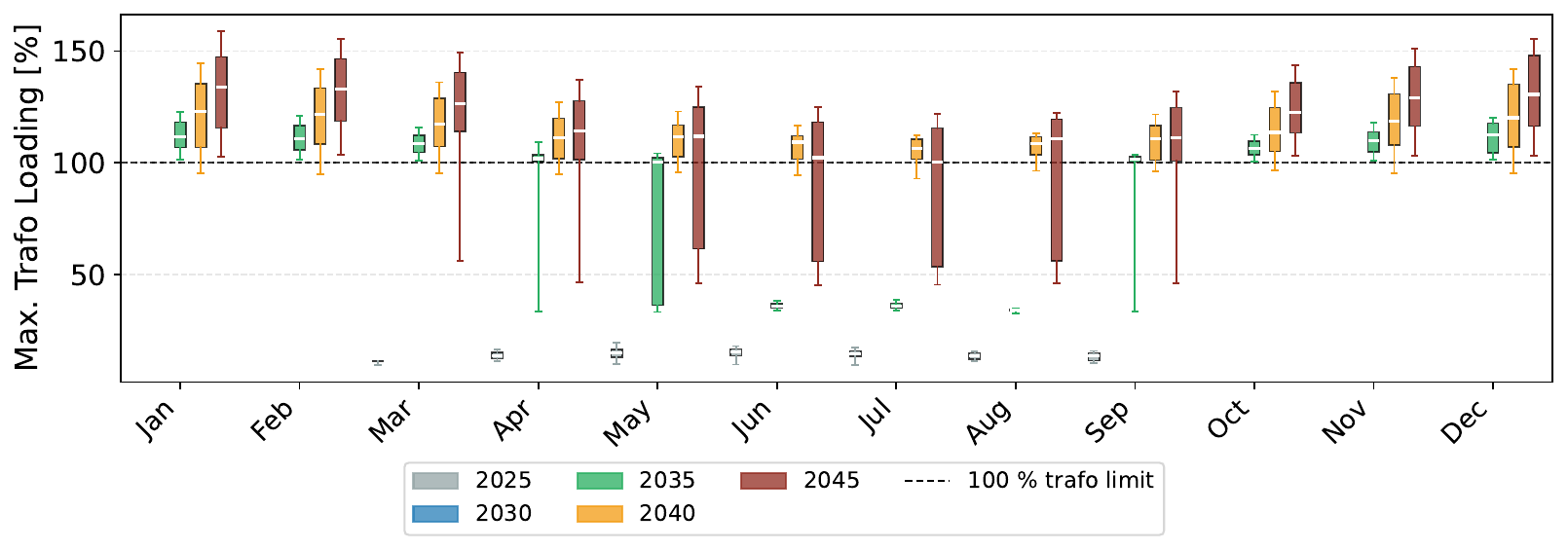}
  \caption{Monthly distribution of transformer loading during congestion periods, poor rural network (2025–2045). Whiskers span the 5\textsuperscript{th}–95\textsuperscript{th} percentile. Dashed line marks the 100\,\% thermal limit. Missing boxes indicate months without congestion events for the respective year.}
  \label{fig:Auslastung_Monatlich_schlecht_laendlich}
\end{figure*}

\begin{figure*}[tp]
  \centering
  \includegraphics[width=0.95\linewidth]{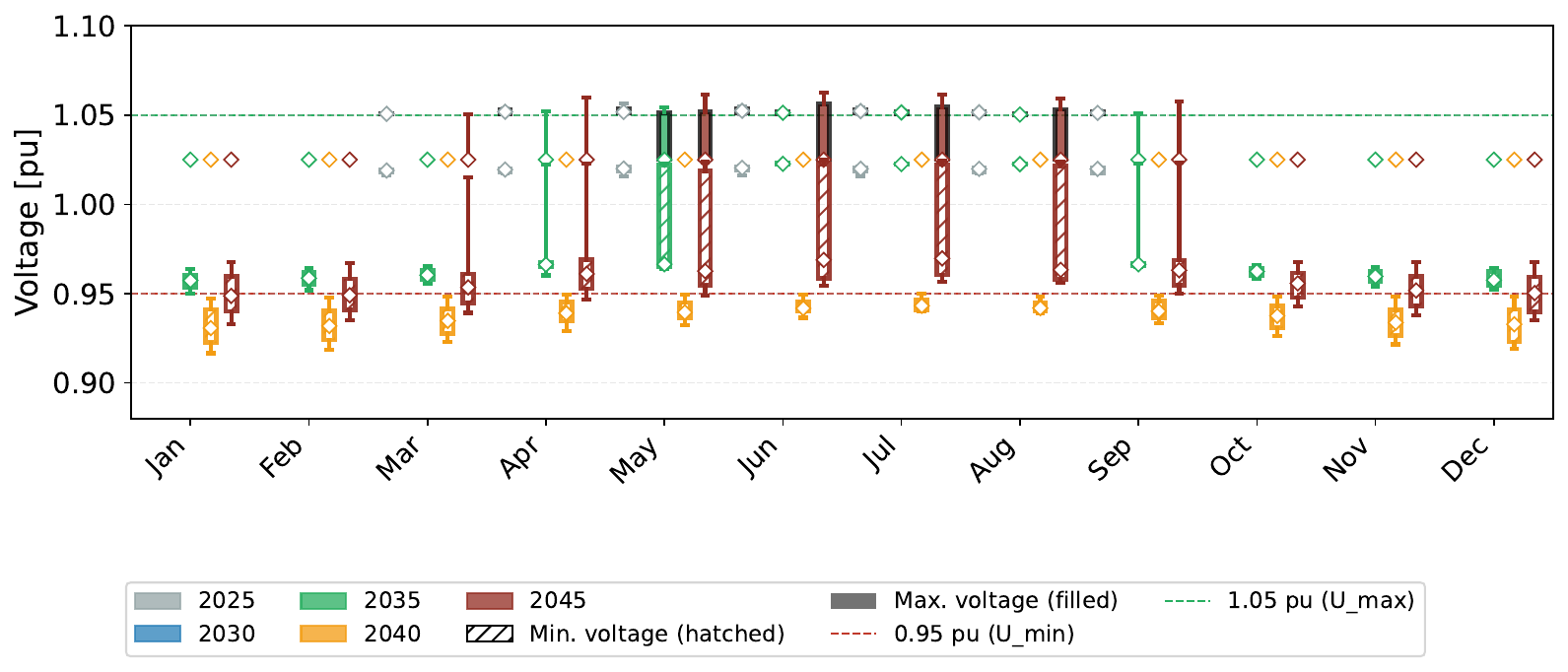}
  \caption{Monthly distribution of bus voltage during congestion periods, poor rural network (2025–2045). Minimum voltage (hatched) and maximum voltage (filled) are overlaid at the same position. By definition, maximum voltage is always above minimum voltage. Where boxes appear very compact, diamond markers indicate the median. Whiskers span the 5\textsuperscript{th}–95\textsuperscript{th} percentile. Dashed lines mark the 0.95\,\text{p.u.} undervoltage and 1.05\,\text{p.u.} overvoltage limits. Missing boxes indicate months without congestion events for the respective year.}
  \label{fig:Spannung_Monatlich_schlecht_laendlich}
\end{figure*}

\autoref{fig:Auslastung_Monatlich_schlecht_laendlich} and \autoref{fig:Spannung_Monatlich_schlecht_laendlich} show the monthly distribution of transformer loading and bus voltages exclusively during congestion periods for the \emph{poor} rural network.

\autoref{fig:Auslastung_Monatlich_schlecht_laendlich} and \autoref{fig:Spannung_Monatlich_schlecht_laendlich} jointly reveal a pronounced bimodal seasonal structure in which two distinct congestion regimes alternate over the year. The 2030 columns are empty throughout, reflecting the congestion-free year identified in the frequency analysis (see \autoref{subsec:devlopmentcong}).

During winter months (November–March), congestion is predominantly load-driven. In 2025, the rural network has essentially no load-driven winter congestion: January, February, November, and December show no boxes at all, and the only winter-month events are a few generation-side ones in March (the low-loading box) in \autoref{fig:Auslastung_Monatlich_schlecht_laendlich}. From 2035 onward, winter congestion emerges with median transformer loading around 105--115\,\%, rising to about 125--135\,\% by 2045 as \ac{EV} and heat pump penetration increases. Simultaneously, minimum bus voltages (hatched boxes) fall to and below the 0.95\,\text{p.u.} limit as \acp{EV} and heat pumps draw power at remote feeder nodes: their median during congestion drops from about 0.96 p.u. in 2035 to 0.93 to 0.95 p.u. from 2040, while maximum voltages (filled boxes) stay near 1.025 p.u., well below the 1.05 p.u. limit, as PV generation is low. These load-driven evening peaks are confirmed in \autoref{subsubsec:intraday_distribution}.

In summer months (May–August), the congestion character depends on the target year. In 2025, all summer congestion events are generation-driven: transformer loading remains low, around 10 to 20\,\% (\autoref{fig:Auslastung_Monatlich_schlecht_laendlich}), reflecting events in which PV backfeed triggers overvoltage at peripheral nodes while the transformer is only lightly loaded. In 2035, the picture is mixed, with transformer loading spanning a wide range from approximately 30\,\% to above 100\,\%, indicating the coexistence of generation- and load-side events within the same season. By 2040, this transitional pattern has resolved into load-side dominance: median transformer loading during congestion exceeds 100\,\% in every month, including June--August, so that the low-loading generation-side summer cluster still visible in 2035 has all but disappeared (only a handful of low-loading generation events remain). By 2045, the summer is load-dominant, with median transformer loading around 100 to 112\,\% across the summer months while a substantial generation-side cluster reappears at midday (roughly 40\,\% of summer congestion events), as growing \ac{EV} and heat pump demand produces evening congestion even in summer months. Correspondingly, maximum voltage values in \autoref{fig:Spannung_Monatlich_schlecht_laendlich} rise toward and beyond the 1.05\,\text{p.u.} upper limit during generation-side events, while minimum voltages fall toward 0.95\,\text{p.u.} when load-side events occur. The load-side share grows strongly through 2040. In 2045, however, a generation-side summer cluster re-emerges as PV backfeed grows, so the shift toward load dominance is not strictly monotonic.

The transition months April and September/October mark the boundary between the two regimes. In April 2035, the transformer loading distribution in \autoref{fig:Auslastung_Monatlich_schlecht_laendlich} spans from approximately 35\,\% to above 100\,\%, and \autoref{fig:Spannung_Monatlich_schlecht_laendlich} shows both overvoltage (maximum voltage near 1.05\,\text{p.u.}) and a load-side minimum-voltage depression to about 0.96 p.u. (still above the 0.95 p.u. limit) within the same month. By 2040, April has shifted to predominantly load-driven congestion (median loading above 100\,\%), and the generation-side signal has largely disappeared. A similar transition is visible in September, where the generation-side events of 2025 give way to load-driven congestion by 2045. October, by contrast, is load-driven in every year in which it shows congestion (from 2035 onward). Load-driven congestion progressively encroaches on the spring and autumn shoulder months. The generation-dominated summer window narrows through 2040 but partially re-widens in 2045 as PV backfeed grows.

Across the congestion-affected years, the transformer-loading distributions intensify and their winter interquartile range widens (about 25 percentage points in 2040 and 29 in 2045). The summer spread is wide only where both regimes coexist within a month: about 60 to 64 percentage points in every summer month of 2045 (midday generation events at 45 to 55\% alongside evening load events above 110\%), and about 66 percentage points in 2035 but only in the mixed month of May. In 2040, with generation-side congestion largely absent, the summer boxes are narrow (about 9 to 14 percentage points) and sit above the 100\% threshold.

Good equipment shows no congestion in any month. The other three network–equipment combinations differ qualitatively. In the \emph{poor urban} network, congestion is exclusively load-driven: no summer generation cluster appears in either the loading or voltage distributions, and congestion occurs in every month, with transformer loading consistently above 100\,\% and the highest event counts in winter. Maximum voltage deviations remain negligible throughout, as the high baseline load suppresses PV surplus at the network level.

For the \emph{medium} equipment level, congestion is entirely absent in the early target years and emerges only from 2035 onward. When congestion does occur, it is exclusively load-dominated in both rural and urban networks: transformer loading during congestion periods is clustered near and above the 100\,\% threshold without a low-loading summer cluster, while bus voltages remain within their limits (no undervoltage violation occurs under medium equipment, as the minimum voltage stays above 0.95 p.u.). This is consistent with \autoref{fig:Entwicklung_Engpasshaeufigkeit}, confirming that PV-driven overvoltage requires substantial prior exhaustion of transformer and cable headroom, which medium equipment does not reach within the simulation horizon.

\subsubsection{Intraday distribution of congestion periods}
\label{subsubsec:intraday_distribution}

\begin{figure*}[tp]
  \centering
  \includegraphics[width=\linewidth]{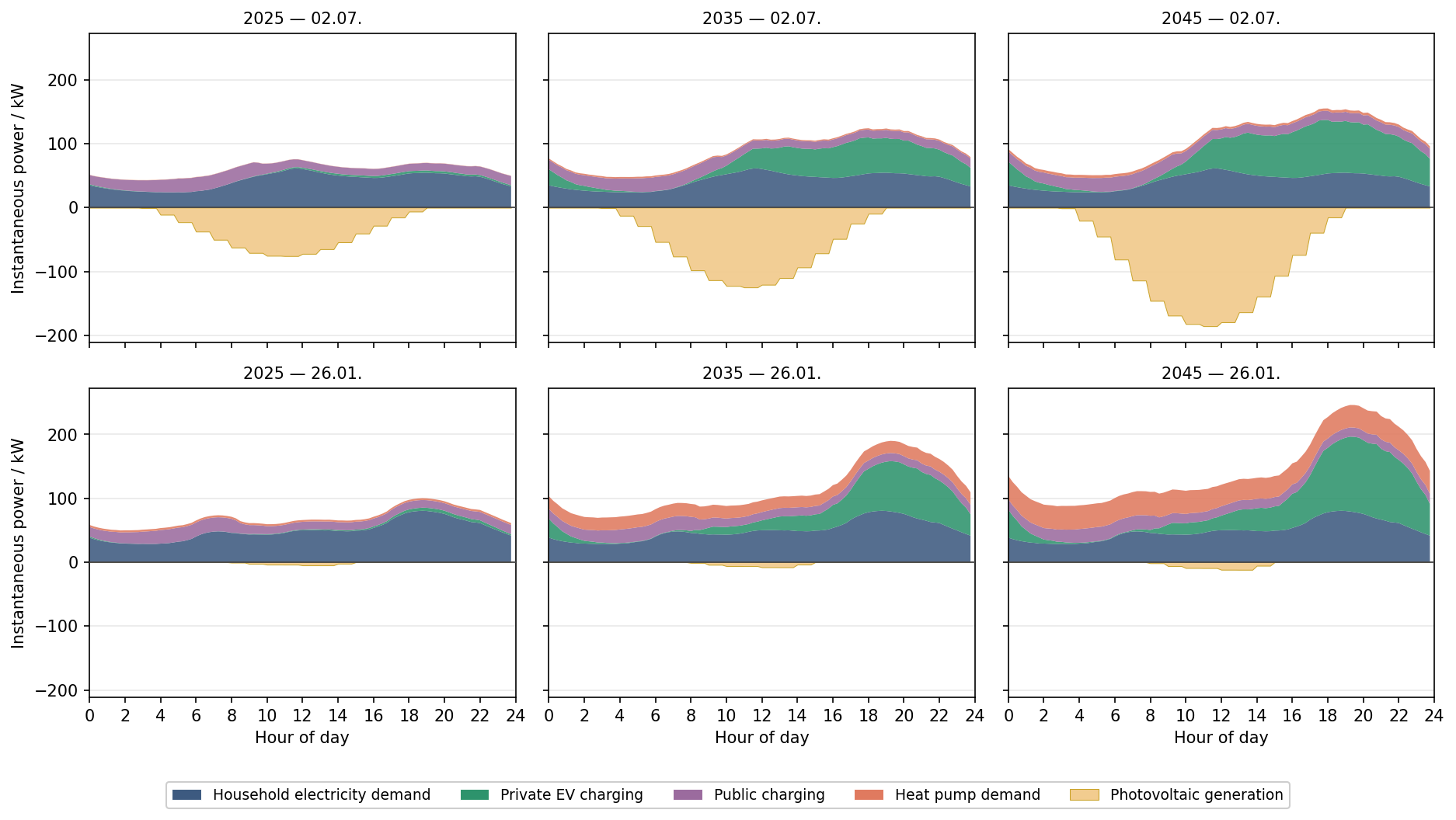}
  \caption{Aggregated instantaneous power for the rural network from 2025 over 2035 to 2045. For each year, the upper panel shows a summer day (02 July), and the lower panel a winter day (26 January). Positive stacks represent household demand, private and public EV charging, and heat-pump demand, while PV generation is plotted below the zero line.}
  \label{fig:stacked_daily_2025_2045}
\end{figure*}

To interpret \emph{when} congestion occurs, hourly congestion statistics must be related to the intraday composition of demand and generation. \autoref{fig:stacked_daily_2025_2045} illustrates the aggregated instantaneous power for the rural scenarios for the years 2025 (current state), 2035, and 2045 (shown as three representative snapshots of the trajectory, with the intervening years 2030 and 2040 lying between them in load and generation magnitude). For each year, the upper panel represents a summer day (2 July) characterized by high PV generation and comparatively low heat pump demand, whereas the lower panel depicts a winter day (26 January) with low PV generation and elevated heat pump demand. Positive stacks disaggregate household electricity demand, private and public EV charging, and heat pump demand, while PV generation is shown below the zero line.
\autoref{fig:stacked_daily_2025_2045} reveals a clear temporal decoupling between midday PV generation and demand components, which peak predominantly in the evening hours. Across the considered years, the positive and negative magnitude increases indicating a progressive intensification of structural network loading.

As EV penetration increases, a pronounced evening charging peak emerges (approximately 18--22\,h) that is only weakly developed in the 2025 baseline. This peak structurally reshapes the daily load profile rather than merely scaling it. As it develops, peak demand shifts into the evening and decouples from the midday PV generation. In the 2025 baseline, peak demand still falls around midday. Summer and winter profiles differ in component magnitude (reduced PV in winter, lower heat pump demand in summer), driven by seasonal solar irradiance and heating demand, respectively.

\begin{figure*}[tp]
  \centering
  \includegraphics[width=\linewidth]{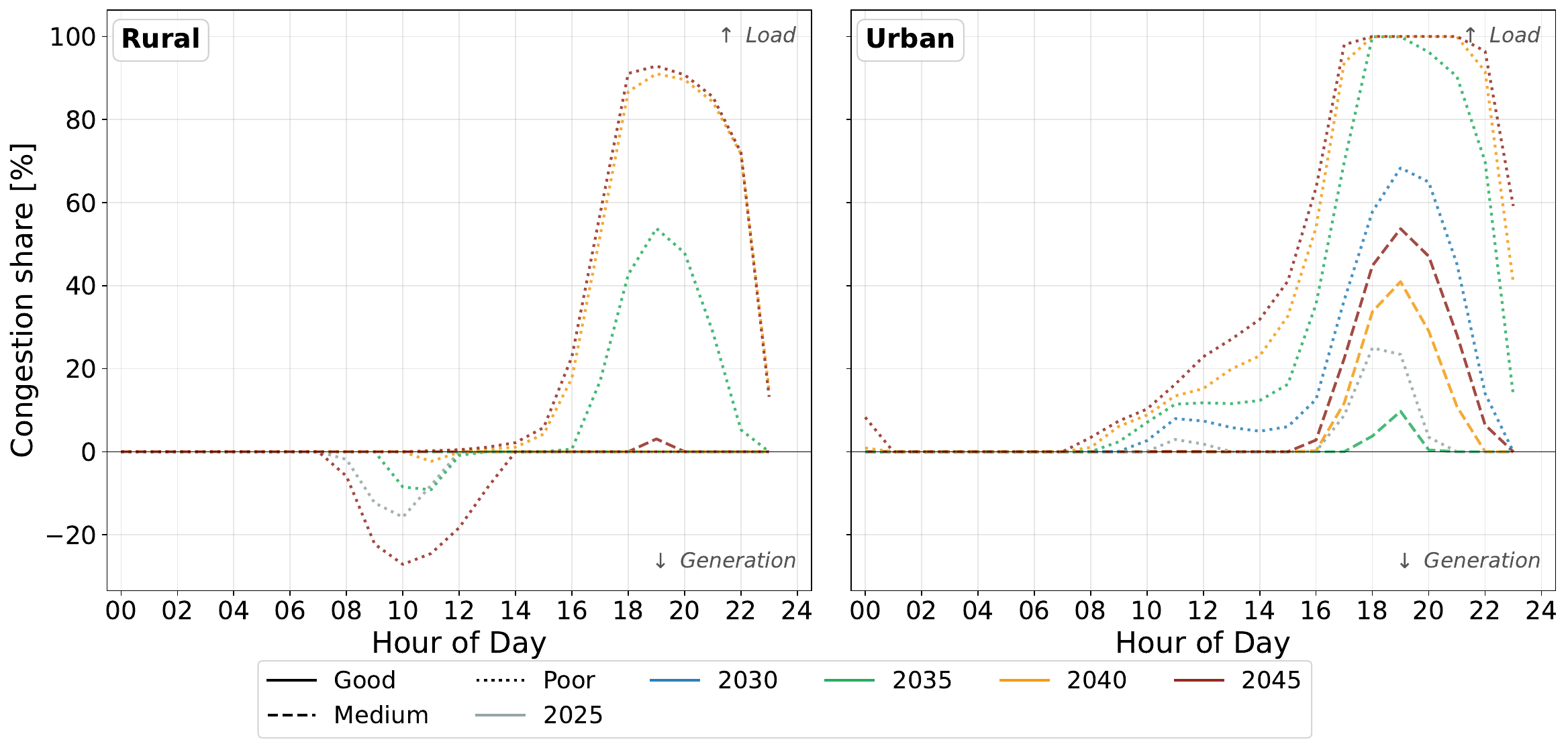}
  \caption{Intraday congestion profiles comparing all three equipment levels in one panel per network type.
           Lines are differentiated by equipment level (solid: good, dashed: medium, dotted: poor)
           and colored by year (2025--2045).
           Each point shows the percentage of quarter-hour time steps at that hour containing at least
           one congestion event.
           Positive values (upward) indicate load-side congestion. Negative values (downward) indicate
           generation-side congestion.
           Lines with zero congestion are omitted.}
  \label{fig:Intrataeglich}
\end{figure*}

Building on this decomposition, congestion periods were aggregated on an hourly
basis and separated into load-side and generation-side events
(\autoref{fig:Intrataeglich}).
A congestion period is classified as generation-dominated when PV generation exceeds
total demand at the transformer level.
Load-side congestion is plotted upward and generation-side congestion downward.

The results reveal that the intraday structure of congestion depends on both the equipment level and the type of congestion. In urban grids, all congestion is exclusively load-driven. Under \emph{poor} equipment in the urban network, load-side congestion concentrates in the late afternoon and evening hours, with peaks between 18:00 and 22:00. This pattern is consistent with simultaneous load peaks from household consumption and, in particular, \acp{EV}. As \ac{DER} penetration increases, the urban congestion window broadens substantially: the evening peak grows from about 25\,\% of time steps in 2025 through 68\,\% in 2030 to 100\,\% from 2035 onward, while its onset shifts from the early evening toward the late morning, until congestion spans almost the entire day by 2045. Under \emph{medium} equipment, a similar but attenuated intraday structure emerges from 2035 onward, predominantly in the urban network (peak congestion share approximately 10\,\%, 41\,\%, and 54\,\% in 2035, 2040, and 2045). Under \emph{good} equipment, no intraday congestion structure is observable.

In rural grids, by contrast, a fundamentally different and bidirectional congestion pattern emerges under poor equipment. Generation-side congestion, driven by PV-induced overvoltage, is concentrated during midday hours (approximately 08:00--14:00), forming a pronounced downward peak that is temporally complementary to the load-side evening peak. Under poor equipment in 2025, generation-side congestion at midday already reaches approximately 15\,\% of time steps, while no load-side congestion is present. This generation-side midday signal is non-monotonic: as a marginal grey-zone signal that sits close to the 1.05 p.u. limit, it recedes just below the limit in 2030, stays minor through 2040, and resurges to its deepest level in 2045. By 2045, generation-side congestion intensifies to approximately 27\,\% during peak PV hours (09:00--13:00), while load-side congestion simultaneously grows to approximately 90\,\% during evening hours (18:00--21:00). This creates two distinct congestion windows per day: a midday generation-side window and an evening load-side window, with a brief transition period around 14:00--16:00 in which congestion of either type is rare.

Under medium and good equipment in rural grids, generation-side congestion is absent or negligible, and the intraday pattern, where congestion occurs, resembles the purely load-driven urban profile.

The bidirectional pattern in poorly equipped rural grids reflects two structurally independent mechanisms: midday PV surplus causes overvoltage at peripheral nodes despite low transformer loading, while the superposition of household, \ac{EV}, and heat pump demand drives evening overloads.

Combined with the winter dominance of load-side congestion (\autoref{subsubsec:seasonal_distribution}), winter evening hours constitute the highest load-side congestion risk, while summer midday hours pose the highest generation-side risk, particularly for poorly equipped rural grids.

\subsubsection{Spatial Distribution and Persistence of Congestion at the Component Level}
The central finding is that the spatial persistence of congestion hotspots is strongly modulated by the equipment level. Under \emph{good} equipment, no congestion corridors form across the entire simulation horizon. Under \emph{medium} equipment, identifiable hotspots emerge from 2035 onward in the urban network, concentrated at the transformer and immediately adjacent feeder sections. Under \emph{poor} equipment, clearly identifiable hotspots appear already in 2025 in both area types, although the rural network is congestion-free in 2030. From 2035 onward the rural hotspots are predominantly load-driven. By 2045, these hotspots intensify and expand spatially.

To quantify this spatial persistence, the frequency with which each individual line and bus was affected by congestion was evaluated across all simulated scenarios. The results confirm a systematic, topology-determined pattern that is most pronounced under poor equipment. Thermal overloads occur at the distribution transformer, while the lines themselves do not experience overloads but show varying degrees of utilization. Aggregated across congestion-affected scenarios and years, the most heavily loaded line segments are located within the first few segments downstream of the transformer in both network types. This reflects the accumulation of downstream power flows at the feeder entry section, where lines carry the combined load of all downstream NCPs. By contrast, voltage limit violations concentrate toward the remote end of the feeders, consistent with the additive voltage drop that accumulates along the feeder path. In the poor rural network this separation is pronounced: the undervoltage violations concentrate at the remote feeder ends (median hop distance about 14 of 27 from the transformer, with the most frequently affected buses near hop 20), whereas the highest-loaded line sits directly at the feeder entry (median hop distance 0 to 1). The structural implication is that both thermally critical nodes and voltage-critical buses are topologically predictable and recurrent. Targeted monitoring can therefore be concentrated at the transformer, the only thermally overloaded element, and at the remote feeder ends, where the undervoltage occurs. The feeder-entry lines carry the highest current but stay within their thermal rating (maximum line loading 70\% rural, 98.6\% urban).

\begin{figure*}[tp]
  \centering
  \subfloat[2025 – mild generation-side event (grey zone): transformer 20\,\%, max.\ line loading 30\,\%, max\,$V=1.057$\,p.u. (12 May, 10:15)]{\includegraphics[width=0.49\linewidth]{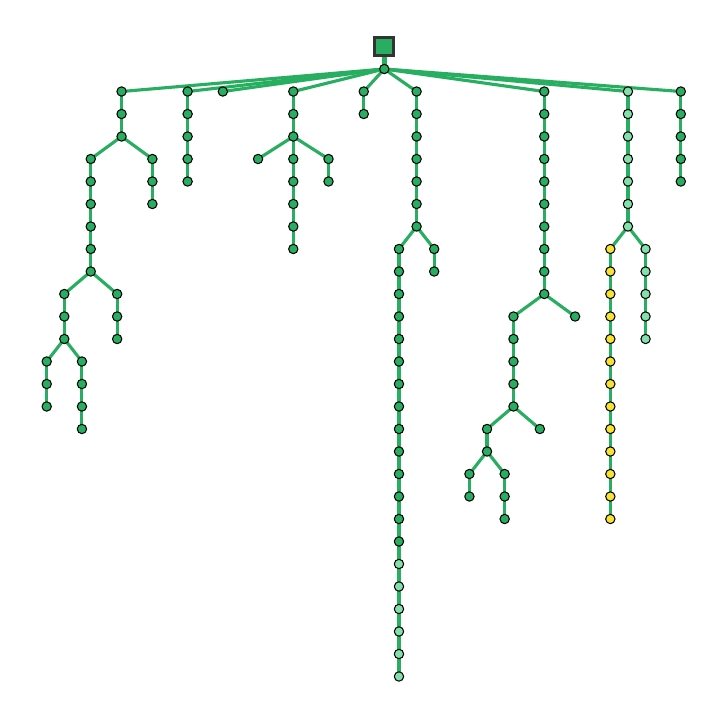}}\hfill
  \subfloat[2045 – severe load-side: transformer 162\,\%, max.\ line loading 70\,\%, min\,$V=0.931$\,p.u. (17 Jan, 19:15)]{\includegraphics[width=0.49\linewidth]{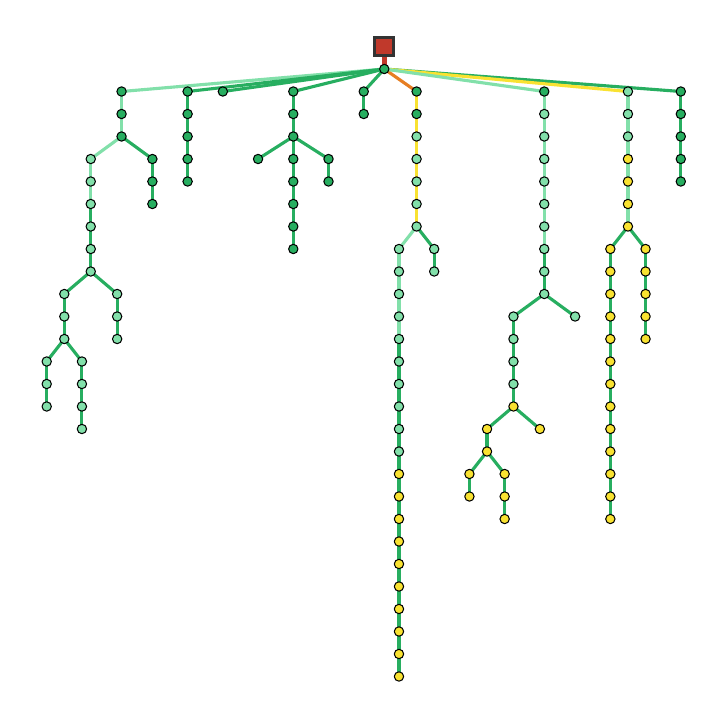}}\\[4pt]
  \includegraphics[width=0.75\linewidth]{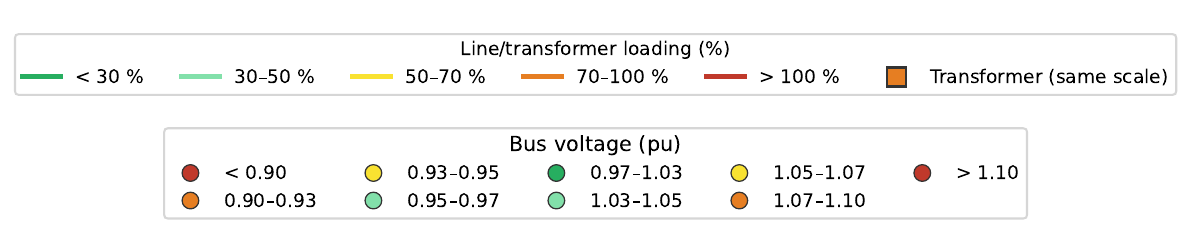}
  \caption{Spatial progression of congestion in the rural SimBench \ac{LV} grid (1-LV-rural3,
           129 buses) under poor equipment. The 2025 panel shows a representative
           generation-side grey-zone event (overvoltage only, no thermal overload).
           The 2045 panel shows the most severe congestion period (highest transformer loading).}
  \label{fig:Netze_example}
\end{figure*}

\autoref{fig:Netze_example} illustrates this progression for the rural network under poor equipment. The 2025 snapshot shows a representative generation-side grey-zone event: at 10:15 on 12 May, PV backfeed exceeds local demand and drives overvoltage at peripheral nodes (max.\ 1.057\,p.u., just above the 1.05\,p.u.\ threshold), while the transformer itself is only lightly loaded (20\,\%) and all line segments remain well below their thermal limits (max.\ 30\,\%). This event is characteristic of the 2025 rural congestion pattern, which consists exclusively of mild PV-driven overvoltage without any thermal overloads. The 2045 panel shows the most severe congestion period, i.e., the time step with the highest transformer loading. By 2045, the worst-case period is load-driven: on 17 January at 19:15, simultaneous \ac{EV} charging and heat pump demand drives the transformer to 162\,\% loading and causes voltage drops at remote feeder ends (min.\ 0.931\,p.u.), while line segments remain below their thermal ratings (max.\ 70\,\%). Under good and medium equipment, this spatial progression is either entirely absent or confined to the transformer vicinity in late target years.

\subsubsection{Combined Temporal Risk Map}

\begin{figure*}[tp]
  \centering
  \includegraphics[width=\linewidth]{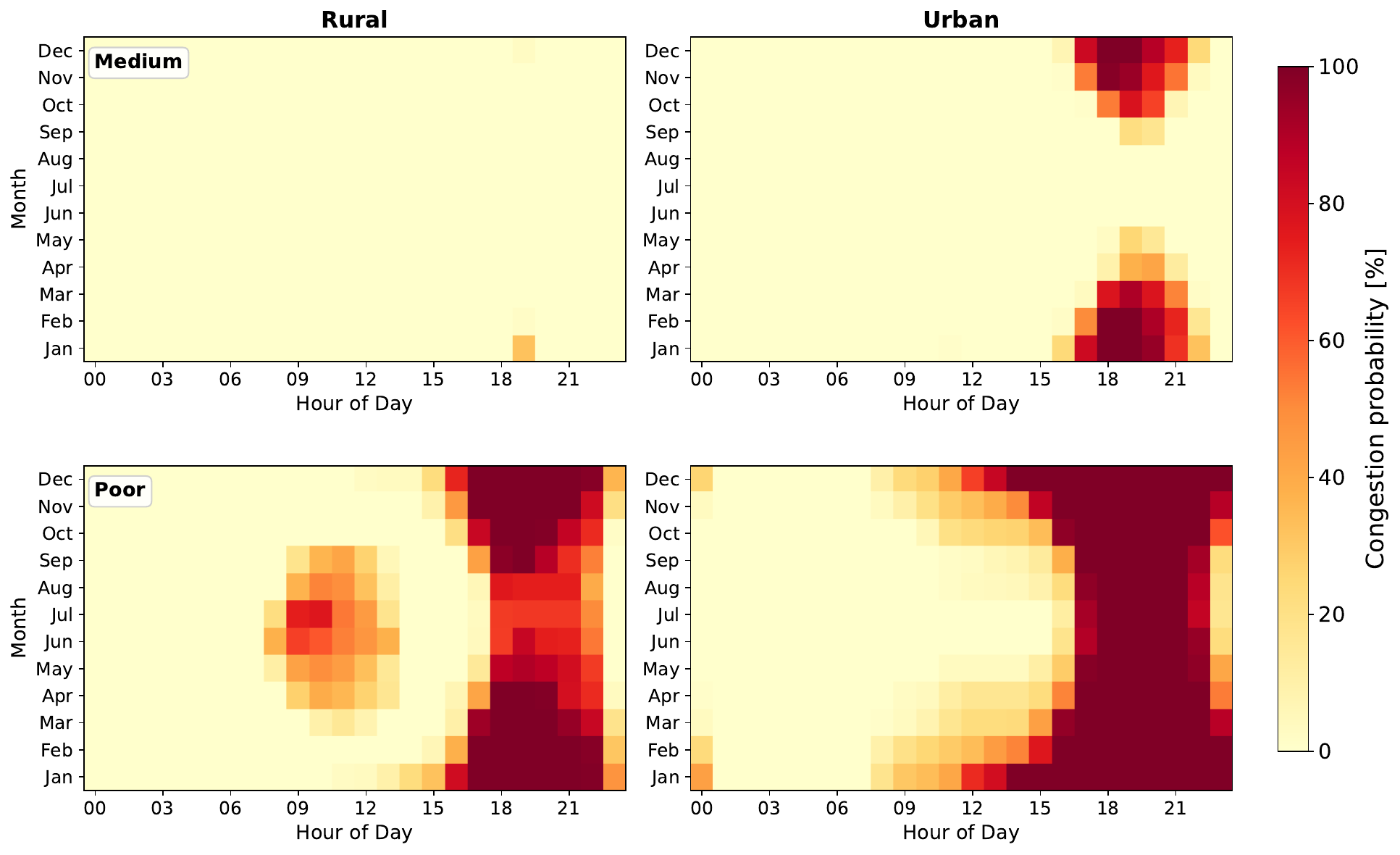}
  \caption{Congestion probability heatmap by month and hour of day for the 2045 target year,
           comparing medium and poor equipment levels (rows) and both network types
           (columns: rural/urban).
           Cell color indicates the fraction of quarter-hour time steps in a given month--hour
           combination that contain at least one congestion event.
           Color scale: light yellow (0\,\%) to red (maximum observed), consistent across all four panels.
           Good equipment is omitted as no congestion events occur.}
  \label{fig:Engpass_Heatmap}
\end{figure*}

\autoref{fig:Engpass_Heatmap} combines seasonal and intraday patterns in a two-dimensional representation, showing the fraction of quarter-hour intervals with congestion for each combination of month and hour of day. 

Under poor equipment, congestion occurs in every month and is concentrated in the afternoon and evening hours (approximately 12:00–24:00). This timing is consistent with the load-driven evening peaks established in \autoref{subsubsec:intraday_distribution}, in particular EV charging. In winter months (November to March) the evening load congestion persists later into the night and, in the urban network, reaches just past midnight (the 00:00 hour is congested in up to 44 percent of winter intervals), while the early-morning hours (01:00 to 07:00) stay congestion-free in both networks. Because the heatmap resolves the seasonal dimension that \autoref{fig:Intrataeglich} averages over, congestion probabilities at late evening hours (21:00–24:00) appear substantially higher in individual winter months than in the annual-average intraday profile, where they are diluted by congestion-free summer months at the same hours.

In the 2045 snapshot, the winter congestion window extends only to the hour just after midnight in the urban network and ends at 23:00 in the rural network, with the early-morning hours staying congestion-free. In summer, the rural network additionally shows the midday generation-side stress identified earlier (peaking near 77 percent around 10:00 in July), whereas the urban network has no generation-side congestion in any month. Summer morning hours remain congestion-free in both.

Under medium equipment, congestion is confined to a narrow evening window (roughly 17:00 to 23:00) and is predominantly urban: it appears from around 2035 in the urban network, whereas the rural network shows only isolated evening congestion from 2045 onward. Good equipment is omitted from \autoref{fig:Engpass_Heatmap}, as no congestion events occur across the entire simulation horizon, consistent with the findings in \autoref{fig:Entwicklung_Engpasshaeufigkeit}.

These differences across equipment levels have direct operational implications: grids with poor equipment require monitoring over a broad temporal range, whereas grids with good equipment remain congestion-free across the simulated scenarios, indicating correspondingly lower monitoring and control needs.

\subsubsection{Structural Synthesis}
The analysis shows that, under the assumption of passive network operation, the grid equipment level is the primary determinant of whether and when congestion occurs, while the stage of DER deployment, together with the settlement type, raises its severity. While well-equipped grids remain congestion-free throughout the horizon, the onset of load-driven congestion is delayed under medium equipment, whereas under poor equipment congestion is present already in the 2025 baseline and reaches high levels by 2045. In 2025 the urban network already shows load-side transformer overload, while the rural baseline is only mild PV-driven overvoltage at low transformer loading.

Increasing \ac{DER} penetration modulates the severity and the seasonal and intraday character of congestion within a given equipment class and can switch congestion on or off over time, as in the poor rural network (congested in 2025, congestion-free in 2030, congested again from 2035). In that case the early generation-side overvoltage is a marginal grey-zone signal that recedes just below its threshold in 2030, before load-driven congestion takes over as deployment continues. The equipment level sets the overall susceptibility to congestion. The following section investigates whether the available measurement infrastructure is sufficient to detect these congestion events via \ac{SE}.

%% file: figures/Ausstattung_NCP_tikz.tex
\definecolor{urbanglow}{RGB}{150,180,235}
\definecolor{urbanedge}{RGB}{70,110,200}
\definecolor{ruralglow}{RGB}{110,205,140}
\definecolor{ruraledge}{RGB}{45,170,90}

\def\ncphouseshape{%
   \fill[black] (-0.40,0.12) -- (0,0.50) -- (0.40,0.12) -- cycle;
   \fill[black] (-0.28,-0.32) rectangle (0.28,0.18);
   \fill[white] (-0.085,-0.32) rectangle (0.085,0.02);}
\def\ncphouse{\begin{scope}[scale=0.62]\ncphouseshape\end{scope}}
\def\ncphhsmall{\begin{scope}[scale=0.52]\ncphouseshape\end{scope}}
\def\ncpcar{\begin{scope}[scale=0.62]
   \fill[black,rounded corners=1.2pt]
     (-0.5,-0.08) -- (-0.5,0.12) -- (-0.28,0.12) -- (-0.16,0.32) -- (0.18,0.32) -- (0.30,0.12) -- (0.5,0.12) -- (0.5,-0.08) -- cycle;
   \fill[black] (-0.27,-0.12) circle (0.11);\fill[white] (-0.27,-0.12) circle (0.045);
   \fill[black] (0.27,-0.12) circle (0.11);\fill[white] (0.27,-0.12) circle (0.045);
   \draw[black,line width=0.7pt] (-0.5,0.02) .. controls (-0.64,0.04) and (-0.64,0.16) .. (-0.62,0.20);
   \fill[black,rounded corners=0.4pt] (-0.72,0.18) rectangle (-0.52,0.32);
   \fill[black] (-0.69,0.32) rectangle (-0.65,0.38);\fill[black] (-0.59,0.32) rectangle (-0.55,0.38);
  \end{scope}}
\def\ncppv{\begin{scope}[scale=0.62]
   \fill[black] (-0.42,-0.02) -- (0.30,-0.02) -- (0.46,0.40) -- (-0.26,0.40) -- cycle;
   \draw[white,line width=0.5pt] (-0.18,-0.02) -- (-0.02,0.40);
   \draw[white,line width=0.5pt] (0.06,-0.02) -- (0.22,0.40);
   \draw[white,line width=0.5pt] (-0.34,0.19) -- (0.38,0.19);
   \draw[black,line width=1.0pt] (-0.06,-0.02) -- (-0.12,-0.28);
   \draw[black,line width=1.0pt] (0.12,-0.02) -- (0.18,-0.28);
   \draw[black,line width=1.0pt] (-0.16,-0.28) -- (0.22,-0.28);
  \end{scope}}
\def\ncphp{\begin{scope}[scale=0.62]
   \fill[black,rounded corners=1pt] (-0.45,-0.28) rectangle (0.45,0.28);
   \fill[white] (-0.14,0) circle (0.18);
   \foreach \a in {30,102,174,246,318}{\draw[black,line width=0.9pt] (-0.14,0) -- ++(\a:0.16);}
   \fill[black] (-0.14,0) circle (0.03);
   \draw[white,line width=0.5pt] (0.15,0.12) -- (0.36,0.12);
   \draw[white,line width=0.5pt] (0.15,0.0) -- (0.36,0.0);
   \draw[white,line width=0.5pt] (0.15,-0.12) -- (0.36,-0.12);
  \end{scope}}
\def\ncpchg{\begin{scope}[scale=0.62]
   \fill[black,rounded corners=1.5pt] (-0.17,-0.34) rectangle (0.17,0.36);
   \fill[white] (-0.115,0.12) rectangle (0.115,0.30);
   \fill[black] (0.04,0.29) -- (-0.06,0.21) -- (-0.005,0.21) -- (-0.04,0.13) -- (0.07,0.22) -- (0.015,0.22) -- cycle;
   \draw[black,line width=0.7pt] (0.17,0.04) .. controls (0.40,-0.02) and (0.34,-0.22) .. (0.30,-0.30);
   \fill[black] (0.265,-0.345) rectangle (0.345,-0.255);
  \end{scope}}

\def\ncpring#1#2#3#4{%
  \begin{scope}[shift={#1}]
    \shade[inner color=#2, outer color=white] (0,0) circle (1.82);
    \fill[white] (0,0) circle (1.04);
    \draw[#3,line width=0.9pt] (0,0) circle (1.06);
    \node[font=\sffamily\small, text=black, align=center] at (0,0) {#4};
  \end{scope}}
\def\ncpframe#1{\begin{scope}[shift={#1}]\draw[rounded corners=12pt, black!75, line width=0.8pt] (-2.7,-2.7) rectangle (2.7,2.7);\end{scope}}
\def\ncpplace#1#2#3#4{\begin{scope}[shift={#1}]\begin{scope}[shift={(#2:#3)}]#4\end{scope}\end{scope}}
\def\ncpleg#1#2#3{\begin{scope}[shift={(#1,-7.0)}]#2\node[font=\sffamily\footnotesize,text=black,anchor=west] at (0.40,0){#3};\end{scope}}

\begin{tikzpicture}
\ncpframe{(-3.6,3.1)}\ncpframe{(3.6,3.1)}\ncpframe{(-3.6,-3.1)}\ncpframe{(3.6,-3.1)}

\ncpring{(-3.6,3.1)}{urbanglow}{urbanedge}{\textbf{NCP} urban\\minimum}
\ncpring{(3.6,3.1)}{ruralglow}{ruraledge}{\textbf{NCP} rural\\minimum}
\ncpring{(-3.6,-3.1)}{urbanglow}{urbanedge}{\textbf{NCP} urban\\maximum}
\ncpring{(3.6,-3.1)}{ruralglow}{ruraledge}{\textbf{NCP} rural\\maximum}

\foreach \a in {30,90,150,210,270,330}{\ncpplace{(-3.6,3.1)}{\a}{1.62}{\ncphhsmall}}

\ncpplace{(3.6,3.1)}{90}{1.62}{\ncphhsmall}

\foreach \a in {45,67.5,90,112.5,135,157.5}{\ncpplace{(-3.6,-3.1)}{\a}{1.62}{\ncphhsmall}}
\ncpplace{(-3.6,-3.1)}{22.5}{2.05}{\ncppv}
\ncpplace{(-3.6,-3.1)}{0}{2.05}{\ncphp}
\ncpplace{(-3.6,-3.1)}{-22.5}{2.05}{\ncpchg}
\ncpplace{(-3.6,-3.1)}{-45}{2.05}{\ncpchg}
\foreach \a in {-67.5,-90,-112.5,-135,-157.5,180}{\ncpplace{(-3.6,-3.1)}{\a}{2.05}{\ncpcar}}

\ncpplace{(3.6,-3.1)}{100}{1.62}{\ncphhsmall}
\ncpplace{(3.6,-3.1)}{38}{2.05}{\ncppv}
\ncpplace{(3.6,-3.1)}{-6}{2.05}{\ncphp}
\ncpplace{(3.6,-3.1)}{-50}{2.05}{\ncpchg}
\foreach \a in {165,210}{\ncpplace{(3.6,-3.1)}{\a}{2.05}{\ncpcar}}

\ncpleg{-6.0}{\ncpcar}{EV}
\ncpleg{-4.0}{\ncphp}{Heat pump}
\ncpleg{-1.5}{\ncphouse}{Household}
\ncpleg{1.1}{\ncpchg}{Charging station}
\ncpleg{5.0}{\ncppv}{PV}
\end{tikzpicture}

%% file: 5_state_estimation_quality.tex
\section{State Estimation Quality Under Varying Measurement Availability} \label{sec:state_estimation_quality}

This section evaluates the quality of \ac{LV} grid \ac{SE} using the \ac{BC-Mod} branch-current \ac{WLS} estimator (\autoref{subsec:bc_mod_estimator}) across the congestion scenarios identified in \autoref{sec:impacts_development_pathways_load_limits}, against the accuracy and detection metrics defined in \autoref{subsec:definition_application_quality_metrics}. The evaluation covers all three equipment quality levels and the three \ac{VDEFNN} 2024 measurement constellations K3 (\ac{SMGW} only), K2 (transformer total power measurement), and K1 (per-feeder measurement). The base scenario set comprises ten combinations (5~years $\times$ 2~network types). Whether a scenario produces congestion depends on the equipment quality level: good equipment yields no congestion in any scenario (0 out of 10), medium equipment produces congestion in 4~scenarios, and poor equipment in 9 of 10 (the 2030 rural network remains congestion-free), resulting in 13~congestion scenarios in total across all equipment levels.

\subsection{Measurement Configurations and Input Data}\label{subsec:data_smgw_penetration_se}

\subsubsection{Measurement Model and VDE~FNN Constellations}

Real measurements represent values acquired by installed \acp{SMGW}
as defined under the German Metering Point Operation Act (Messstellenbetriebsgesetz,
MsbG~\cite{MsbG}). Each real measurement delivers the active and reactive power
injection at a \ac{NCP} for the preceding 15-minute
interval. These measurements are modelled with a relative standard deviation of
$\sigma_\text{real} \approx 1.7\%$ of the measured value~\cite{strobelUncertaintyQuantificationBranchCurrent2024}.
This accounts for both sensor inaccuracy and communication jitter.

The transformer secondary bus (slack bus) serves as the voltage reference node. Its magnitude is
taken from the simulation rather than fixed at nominal: in the evaluated SimBench scenarios the
upstream medium-voltage bus is set to $1.025~\text{p.u.}$ and the on-load tap changer of the
transformer remains at its neutral position, so the secondary reference equals $1.025~\text{p.u.}$
($\widehat{=}~236\,\text{V}$ per phase). Aligning the reference to this value avoids a systematic
offset between the assumed and the actual slack voltage. The slack voltage standard deviation of
$\sigma_\text{slack} = 0.5\%$ accounts for the residual uncertainty in the secondary voltage,
primarily its load-dependent variation across the transformer impedance.

The \ac{VDEFNN} 2024 study~\cite{VDEFNN2024StateEstimationLV} defines three measurement
constellations for \ac{LV} \ac{SE}. All three include \ac{SMGW} real measurements at
equipped nodes and differ in the availability of additional grid-level metering:
\begin{itemize}
  \item K3 – \ac{SMGW} only (no transformer measurement): No transformer or feeder
    measurements are available. Pseudo-measurements for non-metered nodes are derived
    exclusively from standard load profiles scaled by each node's annual energy consumption
    proxy. The \ac{VDEFNN} recommends a minimum \ac{SMGW} penetration of $70\,\%$.
  \item K2 – Transformer total power measurement: In addition to \ac{SMGW} readings,
    the total active power at the transformer secondary is measured. This aggregate value
    is distributed across all non-metered nodes proportionally to their annual energy
    consumption to form pseudo-measurements.
    The \ac{VDEFNN} recommends a minimum \ac{SMGW} penetration of $30\,\%$.
  \item K1 – Feeder-level transformer measurement: In addition to \ac{SMGW} readings,
    each outgoing feeder is individually metered at the transformer secondary.
    Pseudo-measurements for non-metered nodes are generated by distributing the feeder's
    measured active power proportionally to the nodes' annual energy consumption within
    that feeder. The \ac{VDEFNN} recommends a minimum \ac{SMGW} penetration of $15\,\%$
    for radial feeders.
\end{itemize}

Feeder membership is derived algorithmically from the SimBench \ac{LV} network topology
(breadth-first traversal from the transformer \ac{LV} bus), yielding 9 feeders
(rural, 1--29 buses each) and 7 feeders (urban, 1--18 buses each).

\subsubsection{Pseudo-Measurement Model}

Pseudo-measurements follow the proportional load distribution
approach specified in \ac{VDEFNN}~\cite{VDEFNN2024StateEstimationLV, Radhoush2022, Brandalik20171871}.
An estimated or measured total active power is distributed proportionally
to each node's annual energy consumption $E_{a,i}$, obtainable from
metering data:
\begin{equation}\label{eq:weight}
    w_i \;=\; \frac{E_{a,i}}{\displaystyle\sum_{j \in \mathcal{N}} E_{a,j}},
\end{equation}
where $\mathcal{N}$ is the set of all load nodes and $w_i$ is the proportional weight assigned to node $i$. The two measurement-anchored constellations K2 and K1 differ in how the total active power that is subsequently distributed is obtained (K3 must additionally estimate the total load from load profiles, as described below):

\paragraph{K3 (SMGW only)}
Without a transformer measurement, the total grid load must itself be estimated.
Under K3, all non-metered nodes are assigned pseudo-measurements based on a single generic H0 household load profile $f_\text{H0}(t)$ (normalised to unit mean), scaled by each node's consumption proxy:
\begin{equation}\label{eq:k3_pseudo}
    \hat{P}^{(K3)}_i(t) \;=\; f_\text{H0}(t) \cdot w_i \cdot \frac{\sum_{j \in \mathcal{N}} E_{a,j}}{\overline{T}},
\end{equation}
where $\overline{T}$ is the number of hours per year and $\hat{P}^{(K3)}_i(t)$ is the estimated active power consumption of node $i$ at time $t$. In practice this simplifies to
$\hat{P}^{(K3)}_i(t) = f_\text{H0}(t) \cdot \bar{p}_i$, where $\bar{p}_i$ is the average
annual demand rate of node $i$. The H0 profile is a generic household consumption pattern and does not capture the load patterns of \ac{EV} charging, heat pumps, or \ac{PV} feed-in. During congestion periods, which are driven precisely by these devices, the H0-based estimate systematically mis-estimates the true net demand (underestimating load during load-driven periods and missing \ac{PV} feed-in during generation-driven periods). Accordingly, pseudo-measurements under K3 are assigned
$\sigma^{(K3)}_\text{pseudo} = 40\,\%$~\cite{ipachModifiedBranchCurrentBased2021, strobelUncertaintyQuantificationBranchCurrent2024, Angioni2016}.

\paragraph{K2 (transformer total power measurement)}
With the transformer secondary total active power $P_\text{trafo}$ available as a real
measurement, the distribution step replaces the profile-based estimate.
Let $\mathcal{M}$ denote the set of \ac{SMGW}-equipped nodes with real measurements $P^\text{real}_j$.
The residual power after subtracting the known metered injections is distributed
proportionally across the unmetered nodes $i \notin \mathcal{M}$:
\begin{equation}\label{eq:k2_pseudo}
    \hat{P}^{(K2)}_i \;=\;
    \frac{P_\text{trafo} \;-\; \displaystyle\sum_{j \in \mathcal{M}} P^\text{real}_j}
         {\displaystyle\sum_{k \notin \mathcal{M}} w_k}
    \cdot w_i.
\end{equation}
At zero \ac{SMGW} penetration ($\mathcal{M} = \emptyset$) this reduces to
$\hat{P}^{(K2)}_i = P_\text{trafo} \cdot w_i / \textstyle\sum_k w_k$, i.e.,\ proportional
distribution of the full transformer power.
Because the absolute load level is anchored to the measured total, primarily the distributional uncertainty of the consumption-based weights remains (alongside the reactive-power approximation via a constant power factor). This justifies a reduced standard deviation of
$\sigma^{(K2)}_\text{pseudo} = 25\,\%$~\cite{VDEFNN2024StateEstimationLV}.

\paragraph{K1 (feeder-level measurement)}
With per-feeder active power measurements $P_{\text{feeder},f}$ available at the transformer
secondary, the distribution step is performed independently for each feeder $f$.
Let $F_f$ denote the set of all load nodes belonging to feeder $f$, and
$\mathcal{M}_f = \mathcal{M} \cap F_f$ the subset equipped with \acp{SMGW}.
The pseudo-measurement for an unmetered node $i \in F_f \setminus \mathcal{M}$ is:
\begin{equation}\label{eq:k1_pseudo}
    \hat{P}^{(K1)}_i \;=\;
    \frac{P_{\text{feeder},f(i)} \;-\; \displaystyle\sum_{j \in \mathcal{M}_{f(i)}} P^\text{real}_j}
         {\displaystyle\sum_{k \in F_{f(i)} \setminus \mathcal{M}} w_k}
    \cdot w_i,
\end{equation}
where $f(i)$ denotes the feeder of node $i$.
The balance constraint is enforced at feeder level rather than grid level. This confines distributional uncertainty to a smaller group of nodes, which justifies $\sigma^{(K1)}_\text{pseudo} = 15\,\%$~\cite{VDEFNN2024StateEstimationLV}.

\paragraph{Annual consumption proxy}
The proxy $E_{a,i}$ is computed as the sum of all active power load entries at each node, corresponding to metering data available to the \ac{DSO} via the metering point operator. Public charging points are coupled to household buses (see \autoref{sec:impacts_development_pathways_load_limits}) and are implicitly captured in $E_{a,i}$.

\paragraph{Reactive power modeling}
Reactive power is derived from active power via a constant power factor $\cos\varphi = 0.95$. Power values are expressed as single-phase watts, consistent with the \ac{BC-Mod} formulation \cite{strobelUncertaintyQuantificationBranchCurrent2024}.

\subsubsection{SMGW Penetration Configurations}

The fraction of \acp{NCP} equipped with real measurements is referred to as the \emph{measurement
penetration} $n_\text{real} / N_\text{LV}$, where $N_\text{LV}$ denotes the number of
\emph{NCP-occupied} load buses in the \ac{LV} network (excluding the \ac{MV} slack bus and
transformer tap node).
The two SimBench reference networks have $N_\text{LV} = 109$
(rural) and $N_\text{LV} = 53$ (urban).

The evaluation uses the regulatory minimum \ac{SMGW} penetration under the MsbG~\cite{MsbG} (${\approx}\,15\,\%$) as the primary reference point, translating to $n_\text{real}=16$ equipped nodes in the rural network and $n_\text{real}=8$ in the urban network. This level matches the \ac{VDEFNN} minimum for K1 but falls below the recommendations for K2 ($30\,\%$) and K3 ($70\,\%$). The choice reflects the current (2025) rollout status. While \ac{SMGW} penetration will increase under the ongoing \ac{MsbG} rollout, evaluating at this lower bound establishes the worst-case baseline for \ac{SE} accuracy.

Three \ac{SMGW} placement strategies are evaluated:
\emph{power-first} selects the nodes with the highest instantaneous apparent power (oracle upper bound),
\emph{consumption-first} selects by annual energy consumption (feasible \ac{DSO} deployment based on metering data),
and \emph{random} selects uniformly at random (averaged over ten draws to reduce sampling variance). The random strategy simulates an uncoordinated rollout.

The reported accuracy metric throughout this section is the mean normalized voltage error $\bar{\varepsilon}$ (\autoref{eq:norm_voltage_err}, defined in \autoref{subsec:definition_application_quality_metrics}), averaged over the set $\mathcal{T}$ of evaluated congestion periods. This ensures that reported penetration thresholds reflect typical congestion conditions rather than a single extreme snapshot.

To keep the evaluation computationally tractable, up to $|\mathcal{T}| = 200$ congestion periods are selected per scenario. The selection uses stratified sampling to ensure that the sample covers a representative range of operating conditions: periods are binned by transformer loading severity (four quartile bins) and by which line is most heavily loaded (five most-frequent lines plus a residual group), and each bin contributes proportionally to the sample. Scenarios with fewer than 200 congestion periods are evaluated exhaustively.

\subsection{Algorithm Verification}\label{subsec:algorithm_verification_se}

Two correctness tests verify \ac{BC-Mod} under complete real measurement coverage (all load nodes instrumented, $\sigma_\text{real} = 1.7\,\%$), which bypasses pseudo-measurement uncertainty.

A synthetic three-bus network reproduces the reference power-flow voltages with a maximum error of ${<}\,0.0001\,\%$. The same test on the full SimBench rural network ($127$ LV nodes, $30\,\%$ nominal loading) achieves a mean voltage error of $0.0001\,\%$ (max.\ $0.0002\,\%$). This confirms algorithmic correctness. The ground-truth values are obtained from the time-series simulations described in \autoref{sec:impacts_development_pathways_load_limits}.

A mirror test with full measurement coverage (real power and voltage at all nodes) is applied to the first three congestion periods of each base scenario. All scenarios pass the $0.1\,\%$ acceptance criterion (\autoref{tab:mirror_test_results}, \autoref{appendixsection}). The largest deviations (up to $0.09\,\%$, 2045 rural) remain well within the acceptance criterion and reflect residual numerical effects of the \ac{WLS} solution rather than a systematic bias.

\subsection{Estimation Accuracy: K3, K2, and K1 Over All Congestion Scenarios}\label{subsec:simulation_testing_se}

This subsection evaluates the mean normalized voltage error $\bar{\varepsilon}$ (\autoref{eq:norm_voltage_err}) for K3, K2, and K1 across all congestion scenarios using power-first placement and up to 200 stratified congestion periods per scenario.

\begin{figure*}[tp]
  \centering
  \includegraphics[width=\linewidth]{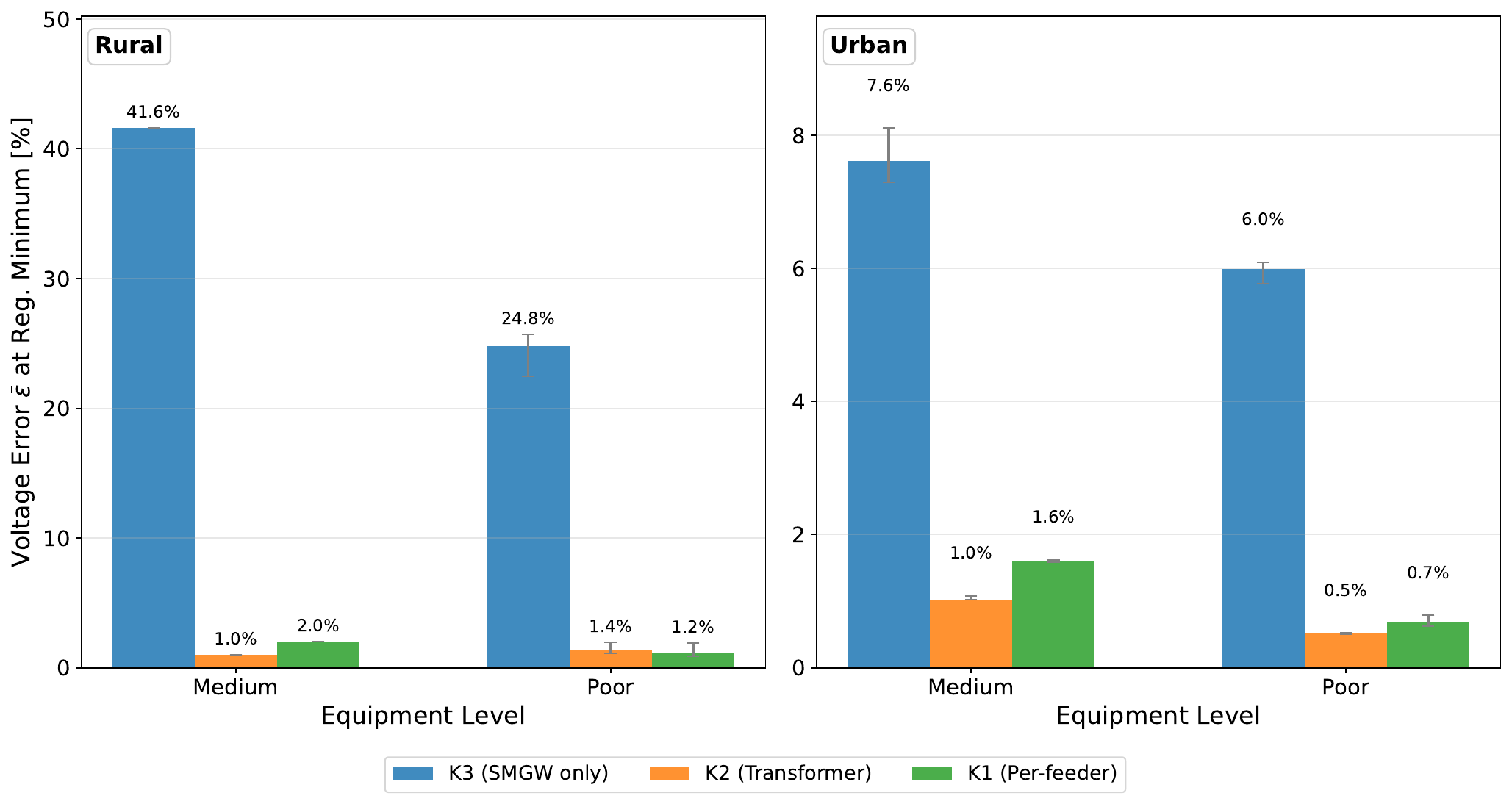}
  \caption{Median voltage error $\bar{\varepsilon}$ at the regulatory minimum (${\approx}\,15\,\%$ penetration, power-first placement)
           by equipment quality level and measurement constellation (K3/K2/K1), for rural (left) and urban (right)
           networks. Bars show the median across congestion scenarios. Error bars span the interquartile range
           (25\textsuperscript{th}--75\textsuperscript{th} percentile).
           Good equipment is omitted as it produces no congestion.}
  \label{fig:SE_Accuracy_Vergleich}
\end{figure*}

\autoref{fig:SE_Accuracy_Vergleich} visualizes the cross-equipment comparison by showing the median voltage error at the regulatory minimum (${\approx}\,15\,\%$) for each equipment level and measurement constellation, separated by network type.

Across all equipment levels, K2 and K1 reduce the median voltage error by roughly an order of magnitude compared to K3, with reductions ranging from about fivefold in the already-low-error urban networks to over fortyfold in the rural networks. Under poor equipment, K2 achieves \mbox{$1.4\,\%$} (rural) and $0.5\,\%$ (urban) at the regulatory minimum, while K3 remains at \mbox{$24.8\,\%$} and $6.0\,\%$ respectively. A single transformer power measurement eliminates the dominant source of pseudo-measurement error (the unknown absolute load level), leaving primarily the distributional uncertainty of the consumption-based weights. This pattern holds under both equipment levels evaluated, although the medium-equipment medians rest on few congestion scenarios (a single rural scenario).

Under K3, the total grid load is estimated from the H0 standard load profile (\autoref{eq:k3_pseudo}), which does not reflect \ac{EV} charging, heat pump, or \ac{PV} feed-in patterns and systematically mis-estimates net demand during congestion periods. Because all pseudo-measurements are derived by distributing this biased total via the weights~$w_i$, installing additional \acp{SMGW} corrects individual nodes but does not fix the systematic total load error. Even increasing \ac{SMGW} penetration to the \ac{VDEFNN}-recommended $70\,\%$ for K3 would not resolve this bias, because the total load estimate remains anchored to the mismatched H0 profile. Under K2, a single transformer measurement eliminates this bias at its source.

The relative performance of K1 and K2 depends on network topology. Under poor equipment, K1 achieves \mbox{$1.2\,\%$} vs.\ K2's \mbox{$1.4\,\%$} in the rural network: the 9 rural feeders (up to 29 buses) carry unevenly distributed loads during congestion, and per-feeder metering captures this spatial heterogeneity. In the urban network, K2 remains slightly better ($0.5\,\%$ vs.\ $0.7\,\%$), as the 7 shorter feeders (up to 18 buses) are more homogeneously loaded, so per-feeder metering adds measurement noise without proportional information gain. Under medium equipment, congestion is milder and the load distribution during congestion periods deviates less from annual averages, so the proportional weights~$w_i$ already approximate the true distribution well. K1's additional feeder constraints provide little benefit in this regime, and K2 outperforms K1 in both networks ($\bar{\varepsilon}$: $1.0\,\%$ vs.\ $2.0\,\%$ rural). Urban networks show lower $\bar{\varepsilon}$ in nearly all configurations due to shorter electrical distances and the resulting flatter voltage profile, which limits error propagation from individual pseudo-measurement inaccuracies.

Under K2 and K1, the three placement strategies (power-first, consumption-first, random) yield nearly identical $\bar{\varepsilon}$ values, because the transformer or feeder-level measurement anchors the total load level and reduces the influence of individual node placement. Under K3, power-first outperforms random placement, but the improvement does not compensate for the systematic H0 bias.

\paragraph{Qualitative illustration}
\label{subsubsec:se_network_illustration}

\begin{figure*}[tp]
  \centering
  \subfloat[Ground Truth\label{fig:se_net_gt}]{\includegraphics[width=0.35\linewidth]{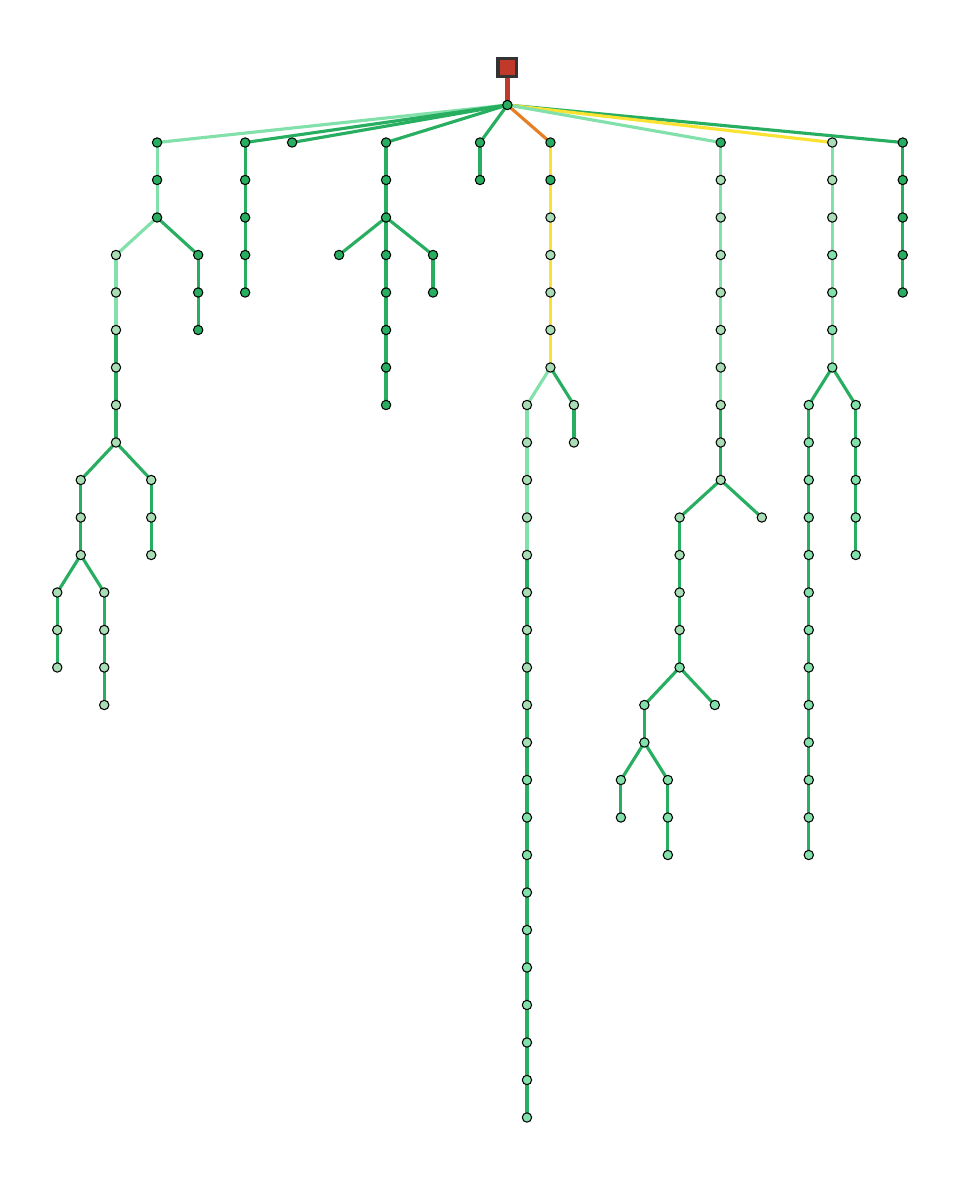}}\hfill
  \subfloat[K3: $\bar{\varepsilon}=39.0\,\%$\label{fig:se_net_k3}]{\includegraphics[width=0.35\linewidth]{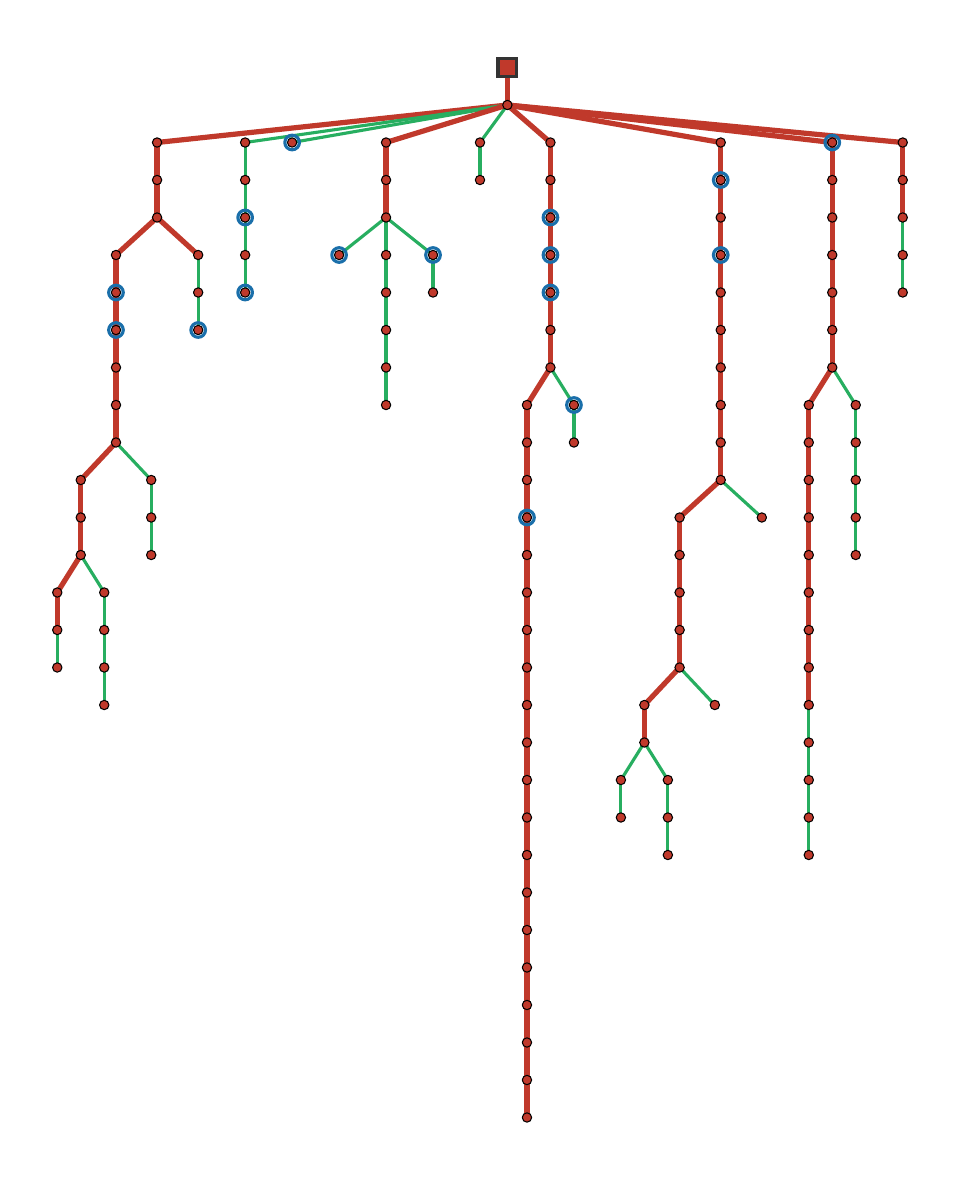}}\\[4pt]
  \subfloat[K2: $\bar{\varepsilon}=2.3\,\%$\label{fig:se_net_k2}]{\includegraphics[width=0.35\linewidth]{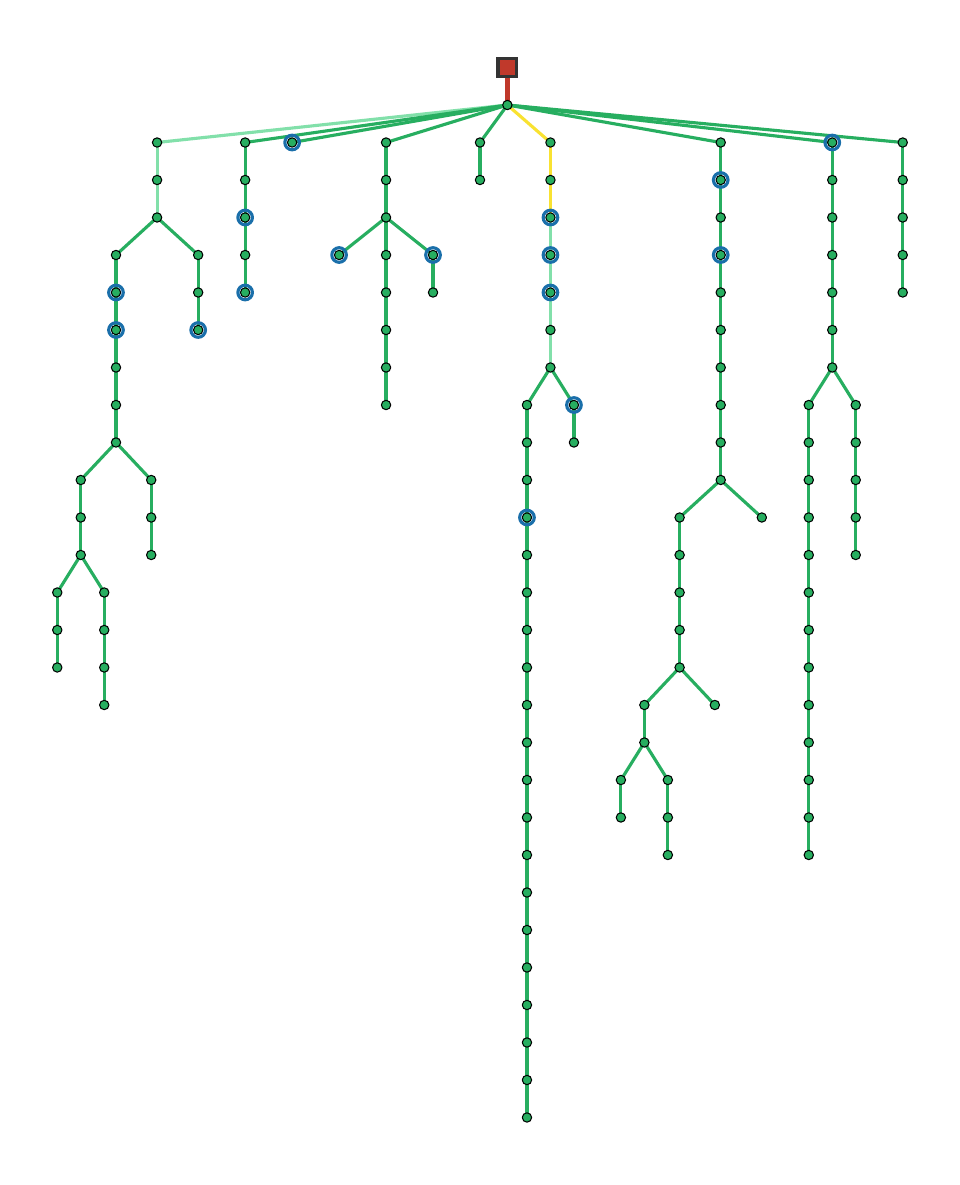}}\hfill
  \subfloat[K1: $\bar{\varepsilon}=1.3\,\%$\label{fig:se_net_k1}]{\includegraphics[width=0.35\linewidth]{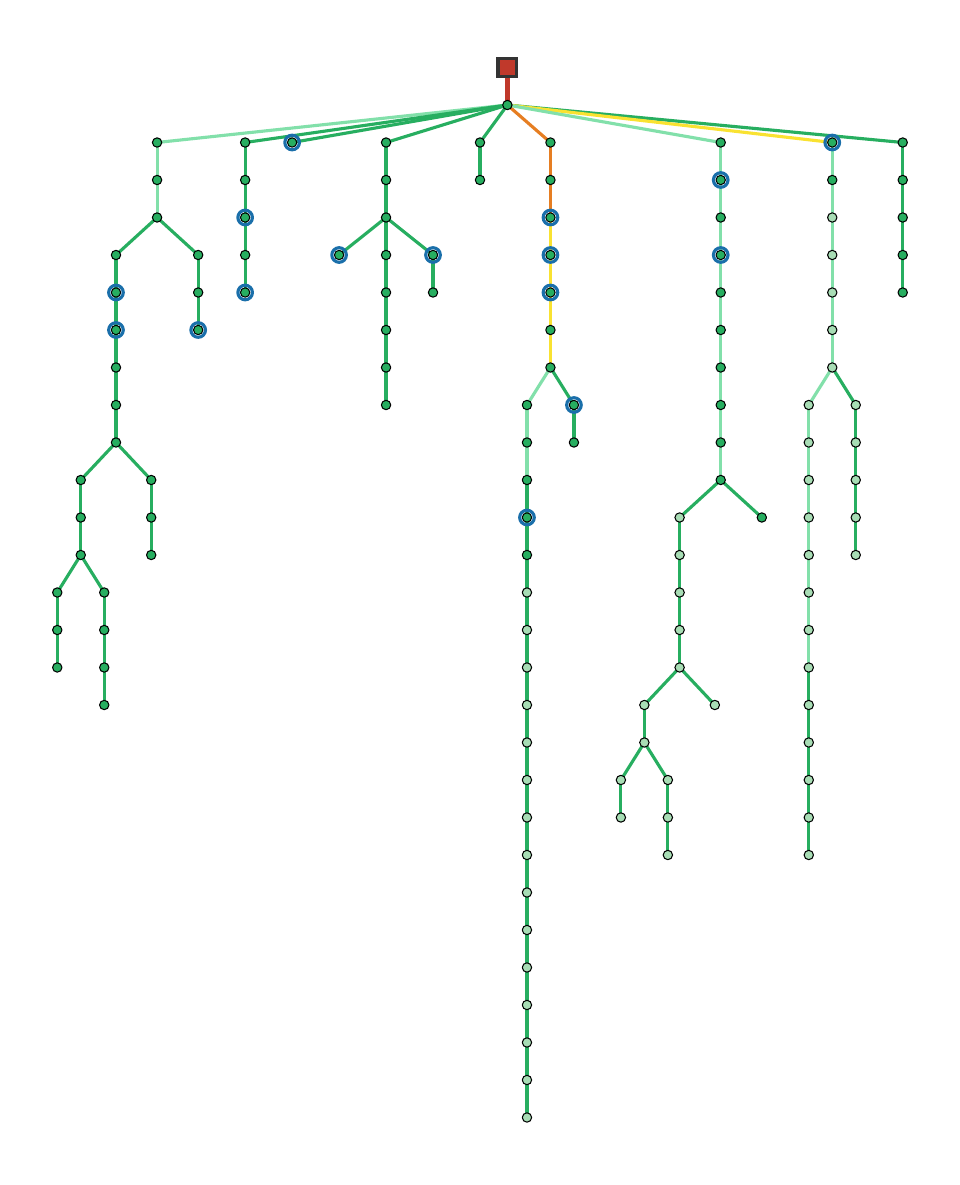}}\\[4pt]
  \includegraphics[width=\linewidth]{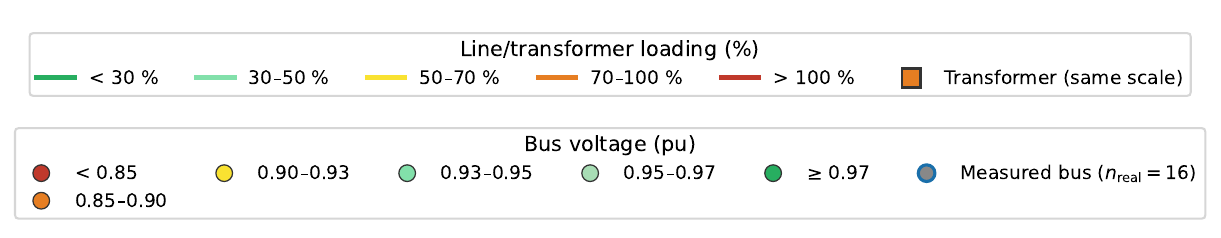}
  \caption{Network state estimation at the regulatory minimum ($n_\text{real}=16$,
           ${\approx}\,15\,\%$ penetration, power-first) for the most heavily loaded congestion period
           in the 2045 rural scenario under poor equipment.
           Subfigure captions give the mean normalized voltage error $\bar{\varepsilon}$.
           Color scales for bus voltage and line loading are shown in the legend.
           Blue ring: bus with real \ac{SMGW} measurement.}
  \label{fig:se_network_k1k2k3}
\end{figure*}

\autoref{fig:se_network_k1k2k3} shows the ground truth and the three state estimates for the most heavily loaded congestion period in the 2045 rural scenario under poor equipment.

K3 (\autoref{fig:se_net_k3}) produces large voltage errors ($\bar{\varepsilon}=39.0\,\%$) and misestimates line currents due to H0 profile mismatch. K2 (\autoref{fig:se_net_k2}) reconstructs voltages accurately ($\bar{\varepsilon}=2.3\,\%$) by anchoring the total load level. K1 (\autoref{fig:se_net_k1}) achieves the lowest error ($\bar{\varepsilon}=1.3\,\%$) through tighter per-feeder constraints.

\subsection{Voltage and Current Accuracy for Congestion Assessment}\label{subsec:congestion_detection_se}

This subsection evaluates \ac{SE} accuracy against \ac{VDEFNN} targets \cite{VDEFNN2024StateEstimationLV} under medium and poor equipment. Congestion is exclusively transformer- and voltage-driven (see \autoref{sec:impacts_development_pathways_load_limits}). No line exceeds its thermal rating (max.\ \mbox{$98.6\,\%$}). Under K2 and K1, the transformer power is directly measured, so the operationally relevant \ac{SE} contribution is the estimation of bus voltages for detecting voltage-band violations. Two \ac{VDEFNN} accuracy metrics\label{subsubsec:detection_metrics} (defined in \autoref{subsec:definition_application_quality_metrics}) are evaluated at the regulatory minimum (${\approx}\,15\,\%$ penetration, power-first): the 99th percentile voltage error $f_V^{p99}$ (target: $\leq 2\,\%$) and the 99th percentile current error $f_I^{p99}$ across lines with ground-truth loading $>20\,\%$ (target: $\leq 10\,\%$)~\cite{VDEFNN2024StateEstimationLV}. Detection performance is assessed through the sensitivity (TPR) and specificity (TNR) defined in the same subsection. Since no line overload occurs, the \ac{VDEFNN} overload TPR cannot be evaluated. The TNR is $100\,\%$ for K2 and K1 (no false alarms), while K3 TNR ranges from $36\,\%$ (poor, rural) to $88\,\%$ (poor, urban).

\subsubsection{Voltage and Current Accuracy at the Regulatory Minimum}
\label{subsubsec:detection_quality}

\begin{figure*}[tp]
  \centering
  \includegraphics[width=\linewidth]{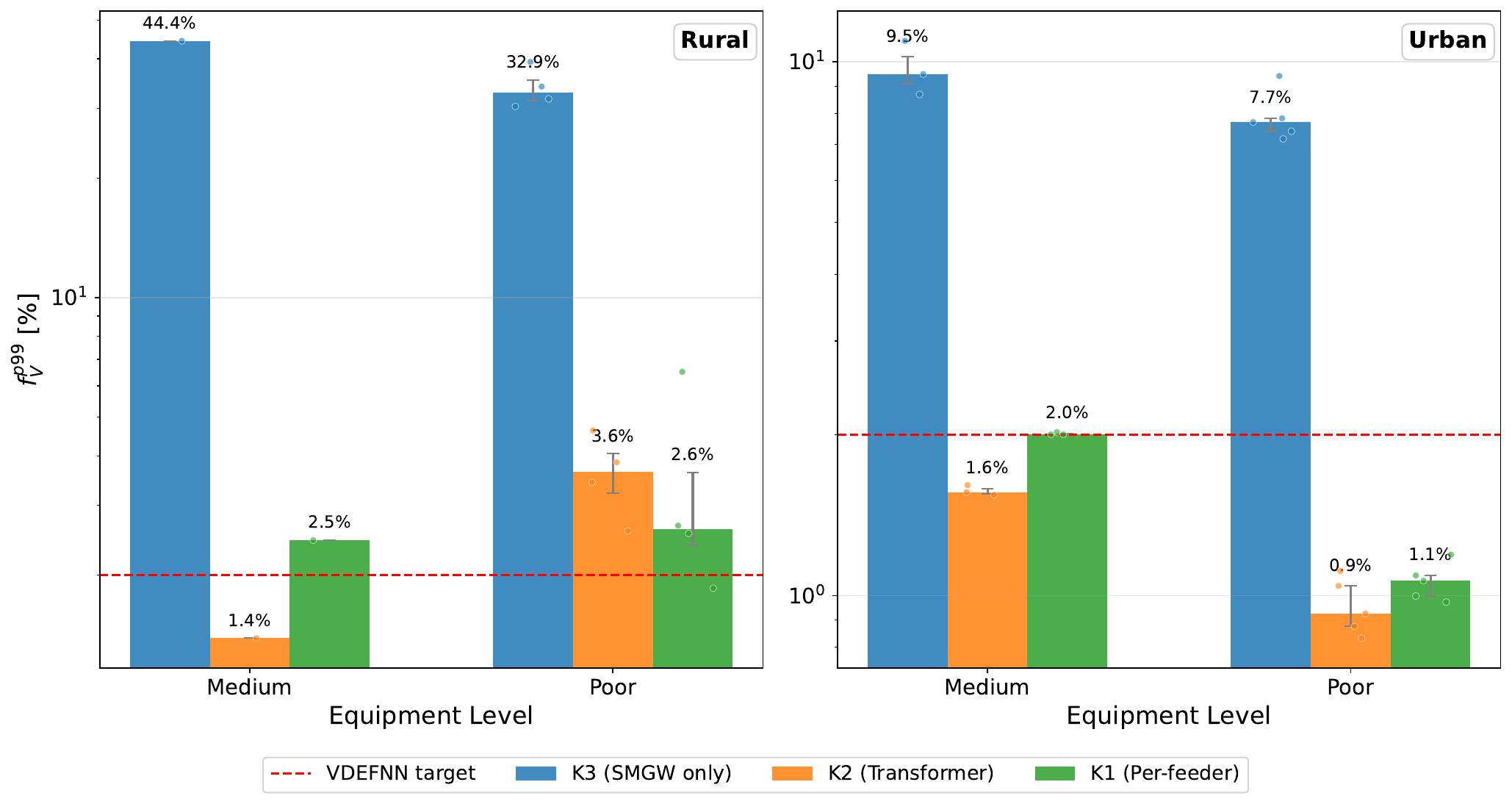}
  \caption{99th percentile voltage error $f_V^{p99}$ at the regulatory minimum (${\approx}\,15\,\%$ penetration, power-first) by equipment quality level and measurement constellation, for rural (left) and urban (right) networks. Bars show the median across congestion scenarios. Error bars span the interquartile range (25\textsuperscript{th}--75\textsuperscript{th} percentile). Dots show individual scenario values. Dashed red line: \ac{VDEFNN} target ($f_V^{p99} \leq 2\,\%$). Good equipment is omitted (no congestion). Note the logarithmic y-axis, different for rural/urban.}
  \label{fig:accuracy_fV}
\end{figure*}

\begin{figure*}[tp]
  \centering
  \includegraphics[width=\linewidth]{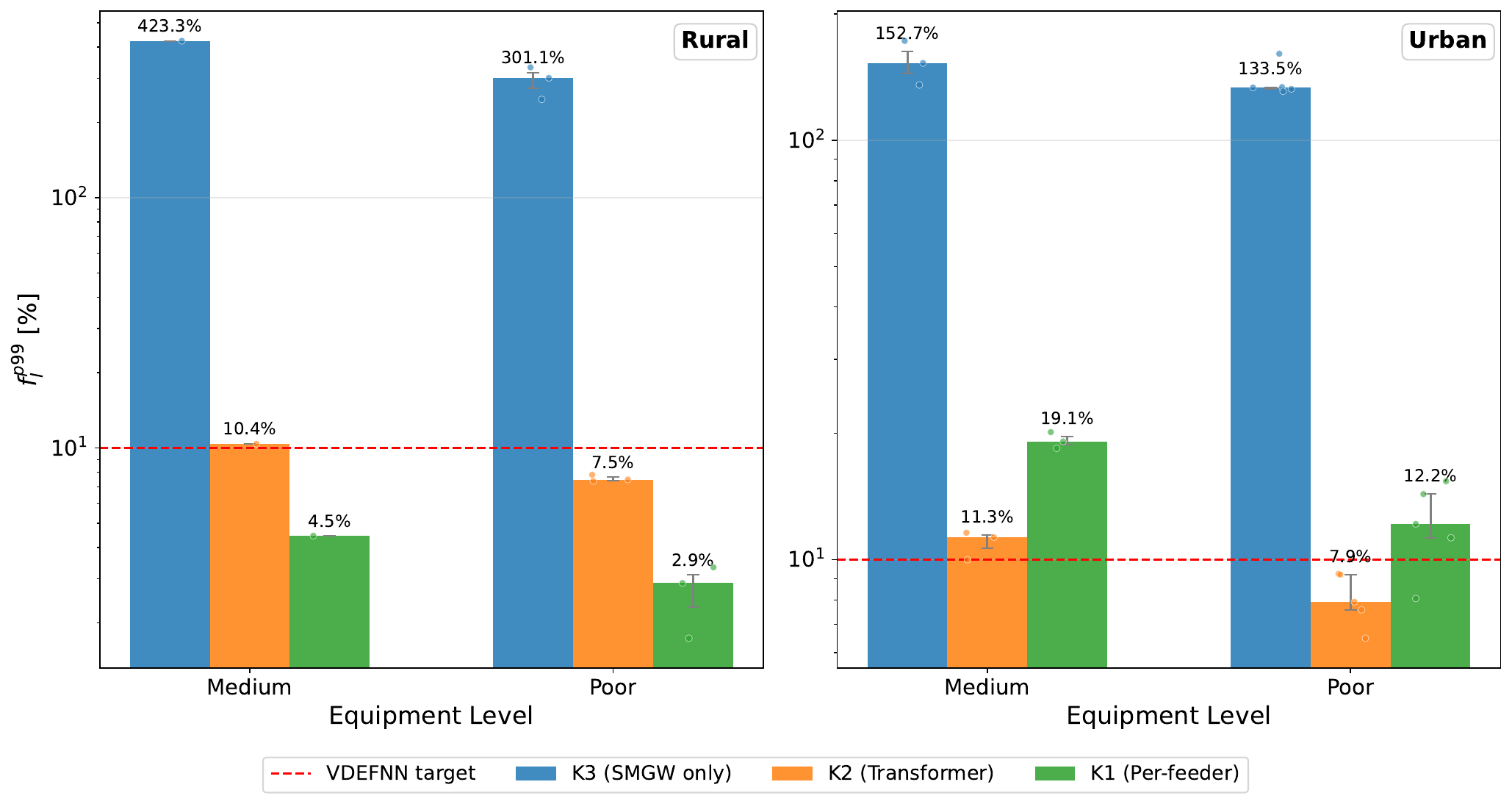}
  \caption{99th percentile current error $f_I^{p99}$ at the regulatory minimum (${\approx}\,15\,\%$ penetration, power-first) by equipment quality level and measurement constellation, for rural (left) and urban (right) networks. Bars show the median. Error bars span the interquartile range (25\textsuperscript{th}--75\textsuperscript{th} percentile). Dots show individual scenario values. Dashed red line: \ac{VDEFNN} target ($f_I^{p99} \leq 10\,\%$). Good equipment is omitted (no congestion). Note the logarithmic y-axis, different for rural/urban.}
  \label{fig:accuracy_fI}
\end{figure*}

\autoref{fig:accuracy_fV} and \autoref{fig:accuracy_fI} visualize the $f_V^{p99}$ and $f_I^{p99}$ distributions at the regulatory minimum across all equipment levels and constellations, with individual scenario values shown as dots where multiple scenarios are available.

\paragraph{Voltage accuracy ($f_V^{p99}$)}
K3 median $f_V^{p99}$ reaches \mbox{$33$--$44\,\%$} (rural) and $8$--$10\,\%$ (urban), far from the $2\,\%$ target and unsuitable for voltage violation detection.

K2 meets the $2\,\%$ target in urban networks ($0.9\,\%$ poor, $1.6\,\%$ medium) and under medium rural equipment ($1.4\,\%$), but reaches \mbox{$3.6\,\%$} under poor rural equipment, which is sufficient for detecting severe violations ($<0.90\,\text{p.u.}$) but not for reliable detection at $0.95\,\text{p.u.}$ The higher error in rural networks is consistent with their longer feeders and larger impedance, which amplify voltage deviations caused by pseudo-measurement errors. Under poor equipment, K1 outperforms K2 in the rural network ($2.6\,\%$ vs.\ \mbox{$3.6\,\%$}), where per-feeder metering better captures the spatially uneven load distribution across feeders, but is slightly worse in the urban network (\mbox{$1.1\,\%$} vs.\ $0.9\,\%$), where the shorter, more homogeneous feeders do not benefit from the additional measurement granularity. Under medium equipment, K2 outperforms K1 in both network types, as the milder congestion conditions produce a more uniform load distribution that the proportional weights already approximate well.

\paragraph{Current accuracy ($f_I^{p99}$)}

K3 current estimation is unusable (median $f_I^{p99}$: \mbox{$133$--$423\,\%$}). K2 meets or approaches the $10\,\%$ target under poor equipment (\mbox{$7.5\,\%$} rural, $7.9\,\%$ urban) and narrowly exceeds it under medium equipment (\mbox{$10.4$--$11.3\,\%$}). K1 achieves \mbox{$2.9\,\%$} in the rural network under poor equipment, where the long feeders carry unevenly distributed currents that per-feeder metering resolves well. In urban networks, K1 reaches \mbox{$12$--$19\,\%$}, because the short feeders provide less current differentiation between nodes, so the per-feeder constraint adds little beyond what K2 already captures.

No constellation consistently meets $f_I^{p99} \leq 10\,\%$ across all configurations, but current accuracy is less operationally relevant since the transformer power is directly measured under K2 and K1.

\paragraph{Implications for voltage-band violation detection}
A violation at $0.95\,\text{p.u.}$ can be reliably detected only if $f_V^{p99}$ is substantially smaller than $5\,\%$. K2 in urban networks ($f_V^{p99} = 0.9$--$1.6\,\%$) and under medium rural equipment ($1.4\,\%$) meets this criterion. Under poor rural equipment ($f_V^{p99} \approx 2.6$--\mbox{$3.6\,\%$}), deep undervoltages ($<0.90\,\text{p.u.}$) are detectable but marginal violations near $0.95\,\text{p.u.}$ may be missed. K3 cannot support voltage violation detection.

\subsubsection{Positioning Relative to State-of-the-Art Methods} \label{subsubsec:se_positioning}

Node-voltage-based and branch-current-based estimators achieve comparable accuracy in the literature. Performance depends primarily on observability and pseudo-measurement quality, not on the choice of state representation~\cite{336098,Primadianto2017}. The K3 errors observed here ($\bar{\varepsilon} \approx 6$--$42\,\%$) are consistent with H0-based estimators reported by \citet{Angioni2016} ($10$--$30\,\%$ at $15\,\%$ penetration) and with the uncertainty bounds in \citet{strobelUncertaintyQuantificationBranchCurrent2024}. K2's median $\bar{\varepsilon}$ of $0.5$--\mbox{$1.4\,\%$} matches state-of-the-art results for estimators with measured transformer injection~\cite{Radhoush2022}. A recent systematic review~\cite{wagnerSystematicReviewState2026} identifies transformer-based anchoring as the single most impactful instrumentation decision for \ac{LV} \ac{SE} accuracy, which matches the findings of the present work.

%% file: 7_discussion.tex
\section{Discussion} \label{sec:discussion}

\subsection{Synthesis of Results}

The combined results of \autoref{sec:impacts_development_pathways_load_limits} and \autoref{sec:state_estimation_quality} reveal a structural mismatch between the pace of \ac{DER}-driven congestion emergence and the measurement infrastructure available for its detection. Grid equipment quality governs congestion onset (0 congestion scenarios under good, 4 under medium, 9 under poor equipment), while the measurement constellation governs whether these congestion events can be detected. A single transformer power measurement (K2) reduces the median voltage error by an order of magnitude compared to K3 and meets the \ac{VDEFNN} voltage target in urban networks. In rural networks under poor equipment, neither K2 nor K1 meets the target, leaving marginal voltage violations near $0.95\,\text{p.u.}$ undetectable. K3 cannot support congestion detection at any \ac{SMGW} penetration level due to the systematic H0-based total load bias.

\paragraph{Transferability}
The analysis is based on two SimBench reference networks (rural and urban) under a single energy transition pathway (Federal Government). While the absolute congestion thresholds and \ac{SE} error magnitudes are network-specific, the structural findings (dominance of transformer overloading and voltage-band violations over line overloads, order-of-magnitude improvement from K3 to K2, K3 stagnation under increasing \ac{SMGW} penetration) are properties of radial \ac{LV} topologies with sparse metering and are expected to hold across networks with similar structural characteristics.

\subsection{Positioning Relative to Existing Regulation and Literature}

\paragraph{§ 14a~\acs{EnWG}}
The German Energy Industry Act, § 14a~\ac{EnWG} \cite{Para14a}, permits \acp{DSO} to curtail controllable loads (\ac{EV} charging stations, heat pumps) in congestion situations in exchange for reduced grid fees. The practical exercise of this right requires the \ac{DSO} to identify congestion reliably and in near-real-time. The present results indicate that, at the current \ac{SMGW} rollout minimum (${\approx}\,15\,\%$ penetration), a \ac{DSO} relying on K3-based \ac{SE} would observe median voltage errors of $6$--$42\,\%$, so reliable congestion identification is impossible. Even under K2, voltage-band violations near the $0.95\,\text{p.u.}$ threshold may go undetected in rural networks where $f_V^{p99}$ exceeds $3\,\%$ (\mbox{$3.6\,\%$} under poor equipment). The § 14a mechanism therefore presupposes a level of grid observability that the current rollout trajectory does not provide in \ac{DER}-dense scenarios.

\paragraph{\acs{MsbG}}
The \ac{MsbG}~\cite{MsbG} defines mandatory \ac{SMGW} rollout for end consumers exceeding specific annual consumption thresholds. While this creates a bottom-up trajectory for \ac{SMGW} deployment, it is decoupled from \ac{DER} expansion: rollout obligations are triggered by consumption, not by congestion risk. A household with an \ac{EV} charger and a heat pump may consume less annual energy than the statutory threshold yet generate disproportionate peak loads that drive transformer overloading. As \ac{DER} penetration increases, more scenarios will require grid monitoring, but the \ac{MsbG} rollout trajectory does not account for this.

\paragraph{\acs{VDEFNN} 2024}
The \ac{VDEFNN} technical specification \cite{VDEFNN2024StateEstimationLV} establishes $f_V^{p99} \leq 2\,\%$ and $f_I^{p99} \leq 10\,\%$ as \ac{SE} quality targets for \ac{LV} grids. This study provides, to the authors' knowledge, the first systematic assessment of their achievability under a concrete \ac{DER} deployment scenario across three equipment quality levels. The results show that the voltage target is met by K2 in urban networks under both equipment levels ($f_V^{p99} = 0.9$--$1.6\,\%$) and in rural networks under medium equipment ($1.4\,\%$). K1 meets the target under poor urban equipment (\mbox{$1.1\,\%$}) but narrowly exceeds it under medium urban equipment ($2.0\,\%$). Neither K1 nor K2 meets the target in rural networks under poor equipment ($2.6$--\mbox{$3.6\,\%$}). K3 is far from the target in all configurations. The current target ($f_I^{p99} \leq 10\,\%$) is met by K2 under poor equipment (\mbox{$7.5$}--$7.9\,\%$) but narrowly exceeded under medium equipment ($10.4$--\mbox{$11.3\,\%$}), and not consistently met by any constellation across all configurations.

The \ac{VDEFNN} study also reports large differences between measurement constellations: the required \ac{SMGW} penetration for radial networks increases from $15\,\%$ with feeder-level measurement to $70\,\%$ with \ac{SMGW}-only setups~\cite{VDEFNN2024StateEstimationLV}. This is qualitatively consistent with the present finding that K3 (no transformer measurement) requires fundamentally higher penetration levels than K2 or K1. The two analyses differ in their evaluation methodology: the \ac{VDEFNN} study restricts accuracy evaluation to the top $1\,\%$ of time steps with highest loading across a full year, whereas the present work evaluates exclusively during congestion periods identified from scenario-based simulations with pathway-specific \ac{DER} deployment. Both approaches focus on high-stress conditions, but the present analysis conditions on actual congestion events rather than a fixed percentile of loading.

The \ac{VDEFNN} specification defines detection metrics (sensitivity and specificity) for three congestion types: transformer overloading, cable overloading, and voltage-band violations~\cite{VDEFNN2024StateEstimationLV}. In the evaluated scenarios, no line exceeds its thermal rating (maximum: \mbox{$98.6\,\%$}), so cable overload detection cannot be evaluated. The dominant congestion mechanisms in these radial \ac{LV} networks are transformer overloading and voltage-band violations, consistent with the congestion types in the \ac{VDEFNN} framework. However, the present results show that the transformer power is directly measured under K2 and K1, so the operationally relevant \ac{SE} contribution is the detection of voltage-band violations.

\paragraph{LV State Estimation Literature}
The \ac{BC-Mod} branch-current formulation belongs to the class of \ac{WLS} state estimators for radial \ac{LV} networks. The mean errors observed under K3 ($\bar{\varepsilon} \approx 6$--$42\,\%$) are consistent with the range reported for profile-based pseudo-measurement approaches: Angioni et al.~\cite{Angioni2016} report mean errors of $10$--$30\,\%$ for H0-based estimators in radial feeders at $15\,\%$ penetration. Koch et al.~\cite{kochSmartMetersGrid2023} show that smart meter substitution becomes feasible only at deployment rates of $30$--$60\,\%$, consistent with the finding that K3 accuracy does not improve meaningfully below these levels. Von der Heyden et al.~\cite{vonderheydenPrivacyPreservingPowerFlow2025} address the complementary problem of data privacy by demonstrating that power flow analysis can be performed on cryptographically hidden prosumer data, which may help overcome regulatory barriers to higher measurement penetration.

A recent systematic review~\cite{wagnerSystematicReviewState2026} emphasizes that
state estimation accuracy in \ac{LV} grids is primarily driven by observability and the joint design of sensing, data fusion, and estimation. The present finding that transformer-based anchoring substantially reduces voltage errors across all equipment levels confirms this.

This work couples \ac{SE} performance to explicit \ac{DER} deployment scenarios across three equipment quality levels. The number of scenarios requiring \ac{SE}-based monitoring grows with increasing electrification, while per-scenario accuracy is primarily determined by the measurement constellation.

\subsection{Derived Regulatory Implications}\label{subsec:regulatory_implications}
The quantitative findings directly motivate three regulatory adjustments to the current German metering and monitoring framework.

\paragraph{Transformer instrumentation as a universally effective first step}
A single transformer power measurement (K2) reduces the median voltage error at the regulatory minimum compared to K3: from \mbox{$24.8\,\%$} to \mbox{$1.4\,\%$} under poor equipment in rural networks, and from $6.0\,\%$ to $0.5\,\%$ in urban networks. In urban networks under both poor and medium equipment, K2 achieves the \ac{VDEFNN} voltage target ($f_V^{p99} \leq 2\,\%$). Under K2 and K1, the transformer power is directly measured, providing immediate congestion detection for the dominant congestion mechanism (transformer overloading). The accuracy improvements (see \autoref{subsec:simulation_testing_se}) confirm that transformer instrumentation provides the largest marginal gain per additional device.

\paragraph{Risk-based supplement to the \acs{MsbG} rollout}
The current \ac{SMGW} rollout minimum (${\approx}\,15\,\%$ penetration) does not provide sufficient voltage accuracy for reliable voltage-band violation detection in rural networks under poor equipment (K2: $f_V^{p99} ={}$ \mbox{$3.6\,\%$}, K1: $2.6\,\%$), although K2 meets the target under medium equipment ($1.4\,\%$). Because the \ac{MsbG} ties rollout obligations to annual consumption thresholds rather than congestion risk, \ac{SMGW} density lags behind measurement requirements as \ac{DER} penetration increases. A risk-based supplement that mandates higher rollout densities in feeders with high \ac{DER} penetration or modeled congestion probability is necessary to close this gap.

\paragraph{Alignment of quality targets with observed congestion mechanisms}
The \ac{VDEFNN} framework defines accuracy targets ($f_V^{p99}$, $f_I^{p99}$) and detection metrics (sensitivity, specificity) for transformer overloading, cable overloading, and voltage-band violations~\cite{VDEFNN2024StateEstimationLV}. In the evaluated scenarios, congestion is caused exclusively by transformer overloading and voltage-band violations and no cable overloads occur. Under K2 and K1, the transformer power is directly measured rather than estimated. The operationally relevant \ac{SE} contribution is therefore the detection of voltage-band violations at the $0.95/1.05\,\text{p.u.}$ thresholds. Detection metrics should be evaluated at these operationally relevant thresholds to ensure that the monitoring system is assessed against the congestion mechanisms that actually occur in radial \ac{LV} networks.

The introduction of three equipment quality levels adds a further dimension. \acp{DSO} operating grids with poor equipment face a trade-off between network reinforcement (replacing undersized transformers and cables to delay congestion onset) and digitalization (deploying measurement infrastructure to detect and manage congestion in the existing grid). Under good equipment, congestion is absent and measurement deployment is less urgent. Under poor equipment, nine of the ten scenarios produce congestion, the 2030 rural network being the only exception, and K2 provides sufficient voltage monitoring in urban networks without any \ac{SMGW} deployment. The optimal strategy for \ac{SMGW} densification depends on the local grid condition, the expected pace of \ac{DER} adoption, and the relative cost of physical reinforcement versus sensor deployment.

\subsection{Limitations}

The evaluation is based on SimBench reference networks (1-LV-rural3, 1-LV-urban6) and \acs{NSC}-based load profiles generated within the same simulation environment. While these networks are designed to be representative of German \ac{LV} grid topologies, absolute results such as congestion thresholds, voltage profiles, and \ac{SE} error magnitudes will differ for specific real-world grids. The two networks are contrasting settlement-specific cases rather than a statistically representative sample of the national stock. The rural network in particular is a single-family area with correspondingly high per-home \ac{EV} intensity, so absolute rural congestion magnitudes represent a high-electrification case rather than a national average. Validation of the reported thresholds against field measurement data from operational \ac{LV} grids remains outstanding.

The pseudo-measurements used in \autoref{sec:state_estimation_quality} derive their consumption weights from the same \acs{NSC} annual energy values that underlie the simulation ground truth. In operational \ac{LV} grids, these weights would be constructed from billing data or standard load profiles, which would introduce additional distributional uncertainty. The qualitative conclusions (K3 insufficiency, K2/K1 superiority) would therefore be reinforced rather than weakened under real-world conditions, as the dominant error source under K3 (the H0-based total load estimate) is independent of the weight accuracy.

The \ac{BC-Mod} algorithm linearizes the branch current equations around nominal voltage. The mirror test (\autoref{subsec:mirror_test_appendix}) confirms that the linearization error remains below $0.1\,\%$ under full measurement coverage, but this error grows under extreme loading conditions where voltages deviate substantially from nominal ($< 0.93\,\text{p.u.}$ in the 2045 rural scenario under poor equipment). The reported \ac{SE} errors therefore include a linearization component that a nonlinear solver would avoid. Nonlinear or hybrid \ac{SE} formulations could improve performance at higher computational cost.

Finally, while the grid-side analysis in \autoref{sec:impacts_development_pathways_load_limits} covers both load-driven and generation-driven congestion, the \ac{SE} evaluation in \autoref{sec:state_estimation_quality} focuses on load-driven congestion periods. Generation-driven congestion (e.g., \ac{PV} reverse power flow causing overvoltage) occurs in the rural network under poor equipment from 2025 onward. The \ac{SE} accuracy under these conditions warrants further investigation, as pseudo-measurement uncertainty may differ when the H0 load profile does not account for generation. No line overloads occur in any scenario (maximum line loading: \mbox{$98.6\,\%$}) and networks with higher cable utilization may exhibit different estimation characteristics.

\subsection{Answers to Research Questions}
The three research questions posed in the introduction can now be answered.

RQ1 asks at what points in time thermal and voltage limit violations emerge under different equipment configurations, how they evolve, and which components are primarily affected. The results show that congestion onset and severity are governed by the interaction between \ac{DER} penetration and equipment quality. Under poor equipment, congestion is present from the 2025 baseline in both network types and intensifies substantially by 2045, with hard congestion events increasingly dominating over grey-zone periods. The rural onset is non-monotonic, however: the early PV-driven midday overvoltage is a marginal grey-zone signal that recedes just below its threshold in 2030, leaving that year congestion-free, before load-driven congestion takes over. Under medium equipment, congestion first appears in 2035 (urban) and 2045 (rural). The dominant congestion mechanisms are transformer overloading and voltage-band violations, with no line overloads in any scenario (maximum line loading: \mbox{$98.6\,\%$}). In rural networks, generation-side congestion driven by \ac{PV} feed-in is visible from 2025 under poor equipment, while urban networks exhibit exclusively load-side congestion.

RQ2 asks how \ac{SE} accuracy and congestion detection capability depend on the type of measurement infrastructure, and which constellation meets quality targets for the identified congestion scenarios. The results show that the measurement constellation (K3/K2/K1) is the dominant factor determining \ac{SE} accuracy at the current regulatory minimum (${\approx}\,15\,\%$ penetration). K3-based \ac{SE} produces median voltage errors of $6$--$42\,\%$ and cannot support congestion detection. K2 reduces errors to $0.5$--\mbox{$1.4\,\%$} (median $\bar{\varepsilon}$) and meets the \ac{VDEFNN} voltage target in urban networks ($f_V^{p99} = 0.9$--$1.6\,\%$). In rural networks under poor equipment, K2 achieves $f_V^{p99} ={}$ \mbox{$3.6\,\%$}, sufficient for detecting severe voltage violations but not marginal ones near the $0.95\,\text{p.u.}$ threshold. K1 provides additional benefit in rural networks under poor equipment ($f_V^{p99} = 2.6\,\%$) but does not outperform K2 in urban networks.

RQ3 asks which regulatory adjustments are necessary to close the observability gap. The results motivate three adjustments (detailed in \autoref{subsec:regulatory_implications}). First, transformer instrumentation should be mandated as a minimum requirement, since K2 provides the largest marginal gain in observability per additional device. Second, the consumption-based \ac{MsbG} rollout should be supplemented with risk-based criteria tied to \ac{DER} penetration and congestion probability. Third, \ac{VDEFNN} detection metrics should be evaluated at the $0.95/1.05\,\text{p.u.}$ voltage thresholds used in congestion definitions, so that the monitoring system is assessed against the violations that actually trigger operational responses.

%% file: 8_summary_outlook.tex
\section{Conclusion} \label{sec:conclusion}

This work systematically couples the German Federal Government energy transition pathway (2025--2045) with time-series simulations of representative rural and urban \ac{LV} reference networks across three equipment quality levels, and evaluates the accuracy of \ac{BC-Mod} \ac{WLS} state estimation under the three \ac{VDEFNN} measurement constellations (K3, K2, K1) at the current regulatory minimum \ac{SMGW} penetration.

The grid-side analysis shows that congestion onset and severity are governed by the interaction between \ac{DER} penetration and equipment quality. Under poor equipment, congestion occurs from 2025 onward and intensifies substantially by 2045. Congestion is caused exclusively by transformer overloading and voltage-band violations, with no line overloads (maximum line loading: \mbox{$98.6\,\%$}). The dominant congestion mechanism shifts from generation-side overvoltage (rural, 2025) to load-side transformer overloading (both networks, 2035--2045) as \ac{EV} and heat pump penetration increases.

The \ac{SE} evaluation reveals that the measurement constellation determines estimation accuracy far more than \ac{SMGW} penetration at the current regulatory minimum (${\approx}\,15\,\%$ penetration). K3-based estimation produces median voltage errors of $6$--$42\,\%$ and stagnates regardless of smart meter density due to the systematic H0-based total load bias (see \autoref{subsec:simulation_testing_se}). A single transformer power measurement (K2) reduces the median voltage error by an order of magnitude (\mbox{$1.4\,\%$} vs.\ \mbox{$24.8\,\%$} in rural networks under poor equipment) and meets the \ac{VDEFNN} voltage accuracy target ($f_V^{p99} \leq 2\,\%$) in urban networks under both poor and medium equipment ($0.9$--$1.6\,\%$). K1 meets the target under poor urban equipment (\mbox{$1.1\,\%$}) but narrowly exceeds it under medium urban equipment ($2.0\,\%$). In rural networks under poor equipment, neither K2 nor K1 meets the target ($f_V^{p99} = 2.6$--\mbox{$3.6\,\%$}). Since the transformer power is directly measured under K2 and K1, the operationally relevant \ac{SE} contribution is the detection of voltage-band violations.

Three regulatory implications follow from these findings. First, transformer instrumentation should be prioritized as it provides the largest marginal gain in grid observability per additional device. Second, the consumption-based \ac{MsbG} rollout should be supplemented with risk-based criteria tied to \ac{DER} penetration and local congestion probability. Third, \ac{VDEFNN} detection metrics should be evaluated at the $0.95/1.05\,\text{p.u.}$ voltage thresholds used in congestion definitions, so that the monitoring system is assessed against the violations that actually trigger operational responses in radial \ac{LV} networks.

While the grid-side analysis covers both load-driven and generation-driven congestion, the \ac{SE} evaluation focuses on load-driven congestion periods. Extending the \ac{SE} assessment to generation-dominant periods, where pseudo-measurement uncertainty may differ due to \ac{PV} reverse power flow, remains open. Further priorities include validating the derived thresholds on real feeder data from multiple \acp{DSO}, developing device-type-aware pseudo-measurement models that use smart-meter metadata~\cite{penaherreraEvaluationEnergyDemands2026}, and incorporating nonlinear \ac{SE} formulations to improve accuracy under extreme loading conditions where the \ac{BC-Mod} linearization introduces systematic error.

%% file: acronyms.tex
\acro{cop}[COP]{coefficient of performance}
\acro{DSO}{distribution system operator}
\acro{EV}{electric vehicle}
\acro{NCP}{network connection point}
\acro{DER}{distributed energy resource}
\acro{HH}{households}
\acro{LV}{low voltage}
\acro{SMGW}{smart meter gateway}
\acro{MV}{medium voltage}
\acro{HV}{high voltage}
\acro{PV}{photovoltaics}
\acro{SE}{state estimation}
\acro{WLS}{weighted least squares}
\acro{AC}{alternating current}
\acro{KVS}{distribution cabinet / switching point}
\acro{PM}{pseudo-measurement}
\acro{BC-Mod}{branch-current-based model}
\acro{UQ}{uncertainty quantification}
\acro{MsbG}{Messstellenbetriebsgesetz}
\acro{EnWG}{Energiewirtschaftsgesetz}
\acro{VDEFNN}[VDE~FNN]{VDE Forum Netztechnik/Netzbetrieb}
\acro{iMSys}{intelligentes Messsystem}
\acro{SGIM}{smart grid interface module}
\acro{NSC}{normalized service curve}
\acro{OLTC}{on-load tap changer}
\acro{LCT}{low-carbon technology}
\acro{FCR}{frequency containment reserve}
\acro{p.u.}{per unit}
\acro{ZIP}{constant impedance, constant current and constant power}
\acro{aFRR}{automatic frequency restoration reserve}
\acro{PMU}{phasor measurement unit}
\acro{RTU}{remote terminal unit}
\acro{LNR}{largest normalized residual}
\acro{SLP}{standard load profile}
\acro{TPR}{true positive rate}
\acro{TNR}{true negative rate}
\acro{IQR}{interquartile range}